\documentclass[aps,prd,groupedaddress,amsmath,amssymb,twocolumn]{revtex4-1}

\usepackage{graphicx}
\usepackage{dcolumn}
\usepackage{bm}
\usepackage{ulem}

\newcommand{\bq}{\begin{equation}}
\newcommand{\eq}{\end{equation}}
\newcommand{\bqa}{\begin{eqnarray}}
\newcommand{\eqa}{\end{eqnarray}}
\newcommand{\nn}{\nonumber \\}

\def\be     {\begin{equation}}
\def\ee     {\end{equation}}
\def\bea        {\begin{eqnarray}}
\def\eea        {\end{eqnarray}}
\def\bnn    {\begin{eqnarray*}}
\def\enn    {\end{eqnarray*}}

\begin{document}

\title{Emergent holographic description for the Kondo effect: Comparison with Bethe ansatz}
\author{Ki-Seok Kim$^{1}$, Suk Bum Chung$^{2,3,4}$, Chanyong Park$^{1,5,6}$ and Jae-Ho Han$^{1,5}$}
\affiliation{$^{1}$Department of Physics, POSTECH, Pohang, Gyeongbuk 37673, Korea \\ $^{2}$Center for Correlated Electron Systems, Institute for Basic Science (IBS), Seoul 08826, Korea \\ $^{3}$Department of Physics and Astronomy, Seoul National University, Seoul 08826, Korea \\ $^{4}$Department of Physics, University of Seoul, Seoul 02504, Korea \\ $^{5}$Asia Pacific Center for Theoretical Physics (APCTP), POSTECH, Pohang, Gyeongbuk 37673, Korea \\ $^{6}$Department of Physics and Photonic Science, Gwangju Institute for Science and Technology, Gwangju 61005, Korea}
\date{\today}

\begin{abstract}
Implementing Wilsonian renormalization group transformations in an iterative way, we develop a non-perturbative field theoretical framework for strongly coupled quantum theories, which takes into account all-loop quantum corrections organized in the $1/N$ expansion. Here, $N$ represents the flavor number of strongly correlated quantum fields. The resulting classical field theory is given by an effective Landau-Ginzburg theory for a local order parameter field, which appears in one-dimensional higher spacetime. We confirm the non-perturbative nature of this field theoretical framework for the Kondo effect. Intriguingly, we show that the recursive Wilsonian renormalization group method can explain non-perturbative thermodynamic properties of an impurity, consistent with Bethe ansatz for the whole temperature region.

\end{abstract}

\maketitle

\section{Introduction}

Non-Fermi liquid physics near metallic quantum criticality \cite{SSL_Breakdown_N,Max_Sachdev_SDW}, nature of metal-insulator transitions \cite{MIT}, emergence of exotic quantum liquids in the vicinity of heavy-fermion quantum criticality \cite{HFQCP_Review_I,HFQCP_Review_II}, and physics of rare events in strongly disordered systems \cite{Q_Griffiths} are all beyond the perturbative theoretical framework. Simply speaking, we do not have a theoretical framework on how to calculate correlation functions for these problems. The AdS$_{d+2}$/CFT$_{d+1}$ duality conjecture \cite{AdS_CFT_Original,AdS_CFT_Follow,AdS_CFT_Correspondence,Bianchi:2001kw,deBoer:1999tgo,Verlinde:1999xm,AdS_CFT_Review} with a spatial dimension $d$ claims to solve these problems in a non-perturbative way. We may translate this duality conjecture in the version of condensed matter physics as follows: Putting Landau-Ginzburg effective field theories on curved spacetime manifolds with an extra dimension and solving classical equations of motion for order parameter fields with Einstein's equations of motion for metric components, one can find not only ground states but also correlation functions non-perturbatively.

This remarkable conjecture appealed to both high-energy and condensed matter physics communities \cite{Nakayama:2013is}. Superconductivity in strongly correlated electrons \cite{Superconductivity_Holography}, the Kondo effect \cite{Benincasa:2012wu,Erdmenger:2013dpa,OBannon:2015cqy,Erdmenger:2016jjg}, non-Fermi liquids \cite{NFL_Holography}, fractional quantum Hall phases \cite{FQHE_Holography}, and metal-insulator transitions \cite{MIT_Holography} have been discussed within the holographic duality conjecture. Even experimental data have been compared with theoretical results of this non-perturbative framework \cite{Graphene_AdS_Description}. However, it is completely unknown the connection between ultraviolet (UV) degrees of freedom in strongly coupled quantum field theories and infrared (IR) emergent fields in weakly correlated classical field theories on curved spacetimes with an extra dimension. In particular, the role of the emergent extra dimension in non-perturbative solutions of strongly coupled quantum field theories remains speculative, resulting in the impression that physical perspectives are unclear in the holographic description.

In this study, we propose one concrete realization of the holographic duality conjecture, implementing Wilsonian renormalization group transformations \cite{Wilson_RG_Review} in a recursive way. In other words, starting from an effective ultraviolet (UV) boundary quantum theory, we derive its corresponding infrared (IR) bulk classical field theory, which appears naturally in one-dimensional higher spacetime. It turns out that the emergent extra dimension can be identified with an energy scale of the renormalization group transformation. We show that an effective bulk equation of motion encodes all-loop quantum corrections through the extra dimension, organized in the $1/N$ expansion \cite{Kondo_1_N_Expansion}, where $N$ represents the flavor number of strongly correlated quantum fields. Applying this recursive Wilsonian renormalization group method to the Kondo problem  \cite{Kondo_Textbook}, we succeed in describing the crossover regime from a weakly correlated local moment fixed point at high temperature to a strongly coupled local Fermi-liquid fixed point at low temperature in a non-perturbative way, where the characteristic energy scale is given by the Kondo temperature. Impurity thermodynamics in our non-perturbative description is qualitatively well matched with the Bethe ansatz for the Kondo effect \cite{Bethe_Ansatz}.

%
%

We would like to point out that our way how to implement Wilsonian renormalization group transformations in a non-perturbative way is parallel to that of S.-S. Lee's approach \cite{From_Field_Theory_To_Emergent_Gravity}: Wilsonian renormalization group transformations give rise to double-trace operators, but such interaction terms are translated into single-trace operators with appropriate order parameter fields through the Hubbard-Stratonovich transformation, where these order parameter fields are classical variables. However, there exist two essential different aspects between these two formulations: First, the previous study takes into account renormalization group transformations in real space while the present study implements them in momentum space. Second, the previous emergent gravity description recovers the result of a mean-field theory for the O(N) vector model in the large$-N$ limit while our non-perturbative field theoretical framework gives rise to resummation of higher-order quantum corrections for the Kondo effect beyond the result of a mean-field theory. Involved with the first issue, the former starts from an insulating UV fixed point, where the kinetic-energy term to describe hopping of electrons is considered as a perturbation at the UV fixed point. On the other hand, we start from a metallic UV fixed point, where interaction terms are taken into account as perturbations at the metallic fixed point. It is certainly easier to see the connection between the field theoretical approach and the emergent gravity formulation since both field theoretical and gravity descriptions are based on the same UV fixed point. This is the reason why the present study could reveal physics of the extra dimension clearly. Associated with the second issue, bulk fields identified with order parameter fields are integrated out, responsible for higher-order quantum corrections. On the other hand, such bulk fields are taken into account as background fields, and thus a mean-field theory is reproduced in the previous approach. To integrate over bulk degrees of freedom is an essential ingredient of our present study beyond all existing investigations. We point out recent developments in the derivation of the holographic duality conjecture from field theoretical perspectives based on how to implement Wilsonian renormalization group transformations \cite{MERA_AdS_CFT,RG_AdS_CFT1,RG_AdS_CFT2,RG_AdS_CFT3,RG_AdS_CFT4,RG_AdS_CFT5,RG_AdS_CFT6,RG_AdS_CFT7,RG_AdS_CFT8,RG_AdS_CFT9,RG_AdS_CFT10,RG_AdS_CFT11, RG_AdS_CFT12,QFT2_GR3_Kim_Park,MERA1,MERA2,MERA_Review,Swingle_MERA_AdS}.

Recently, we proposed an emergent geometric description for a topological phase transition in the Kitaev superconductor model, which allows us to extract out an emergent metric structure \cite{Kitaev_TSC_Holography}. Based on the Ryu$-$Takayanagi formula with such a metric tensor \cite{Ryu_Takayanagi_Formula}, we calculated holographic entanglement entropy. Interestingly, it turns out that this entanglement entropy reproduced the Cardy's formula \cite{Cardy_CFT} perfectly not only at but also near the quantum critical point \cite{Kitaev_TSC_Holography}. 

The present manuscript is organized as follows. We start from a strongly coupled quantum mechanics theory to describe the Kondo problem, introduced in section II. Resorting to the Wilsonian renormalization group analysis, we find an effective classical field theory in the large $N$ limit. Here, $N$ corresponds to the number of flavors, more precisely, the spin degeneracy. It turns out that this novel large $N$ classical theory appears in one dimensional higher spacetime, given by $(1+1)D$ Landau-Ginzburg-type quantum field theory for the hybridization order parameter. The emergence of this large $N$ classical field theory in one dimensional higher spacetime is the holography structure in this study. This is one of the main points in our study, introduced in section III A. Based on this effective classical field theory in $(1+1)D$, we investigate thermodynamic properties of this Kondo problem for the range of whole temperatures. We compare both specific heat and spin susceptibility for the impurity dynamics from our emergent holographic description with Bethe ansatz results, discussed in section III B. An essential point is that our renormalization group procedure seems to be one path integral reformulation for the Wilsonian numerical renormalization group structure, where the whole procedure is discussed in sections IV A, IV B, and IV C. The connection between our path integral formulation and the numerical renormalization group method has been discussed in section IV E. This serious comparison, in spite of ``speculative", reveals the origin for the emergence of the extra dimension. In particular, we demonstrate explicitly that the evolution of the local order parameter through the extra dimension introduces quantum corrections order by order reorganized in the $1/N$ expansion for the boundary quantum mechanics theory. See section IV D. Our path integral reformulation shows that summing up such quantum corrections non-perturbatively in the all-loop order gives rise to a novel large $N$ field theory in the holographic structure. Section V discusses how to calculate the thermodynamics in the Kondo problem from our holographic dual field theory. One important question which remains is how we find a structure of curved spacetime in this effective classical field theory. This curved nature of spacetime turns out to be essential to encode effects of strong correlations in the original conjecture for holography. In sections VI A, VI B, and VI C, we discuss how to extract out an emergent metric structure from this effective classical field theory. Unfortunately, this translation from a field theory to an Einstein equation has not been resolved clearly in the present study. The research direction for geometric translation deserves further serious investigations. Section VII concludes this study with detailed summary.

\section{The Kondo problem: An effective quantum-mechanics theory for the Kondo effect}

The meaning of the Kondo problem is as follows. An impurity spin is weakly correlated with itinerant electrons at high temperatures and thus, the perturbation theory works perfectly well above the Kondo temperature. On the other hand, spin flip scattering become stronger due to the nonabelian nature of effective interactions, decreasing temperature to approach the Kondo temperature. Finally, the effective interaction parameter renormalizes to diverge at the Kondo temperature. As a result, the perturbation theory breaks down below the Kondo temperature. The Kondo problem is how to describe the strong coupling fixed point, starting from the weak coupling fixed point, which is certainly beyond the perturbative theoretical approach.

We start from the path-integral representation for the Kondo problem, given by
\bqa && Z = \int D c_{\sigma}(\bm{k},\tau) D \bm{S}(\tau) \ e^{-S}, \nn
&&S = - S_{B}[\bm{S}(\tau)] \nn && - \int_{0}^{\beta} d \tau \bigg\{ \int \frac{d^{d} \bm{k}}{(2\pi)^{d}} c_{\sigma}^{\dagger}(\bm{k},\tau) \Big( \partial_{\tau} - \mu + \frac{\bm{k}^{2}}{2m} \Big) c_{\sigma}(\bm{k},\tau) \nn && + \frac{J_{K}}{N} \int \frac{d^{d} \bm{k} d^d \bm{k'}}{(2\pi)^{2d}} c_{\alpha}^{\dagger}(\bm{k},\tau) \bm{\sigma}_{\alpha\beta} c_{\beta}(\bm{k}',\tau) \cdot \bm{S}(\tau) \bigg\} . \eqa Here, $c_{\sigma}(\bm{k},\tau)$ is an electron field, where the spin degeneracy is extended from $\sigma = (\uparrow, \downarrow)$ to $\sigma = (\uparrow, \downarrow) \otimes (1, ..., N)$. $\bm{S}(\tau)$ describes an impurity spin, where $S_{B}[\bm{S}(\tau)]$ is its Berry-phase action. $J_{K}$ is the Kondo coupling constant, scaled by $N$ for the $1/N$ expansion. In this paper we use the Einstein convention, where the symbol of spin summation is omitted for simplicity.

In order to describe the Kondo effect and deal with the Berry-phase action, it is conventional to write an impurity spin with the Abrikosov fermion variable in the Sp(N) representation, given by
\bqa && \bm{S}(\tau) = \frac{1}{2} f_{\alpha}^{\dagger}(\tau) \bm{\sigma}_{\alpha\beta} f_{\beta}(\tau) , \eqa
where these fermions should satisfy the single occupancy constraint
\bqa && f_{\sigma}^{\dagger}(\tau) f_{\sigma}(\tau) = N S \eqa
with $N S = 1$ \cite{Kondo_Textbook}. Inserting this expression into the Kondo model, we obtain
\bqa  Z &=& \int D c_{\sigma}(\bm{k},\tau) D f_{\sigma}(\tau) D \lambda(\tau) \ e^{-S}, \nn S &=& \int_0^\beta\!d\tau \bigg\{ \int\!\frac{d^d\bm k}{(2\pi)^d} \ c_\sigma^\dagger(\bm k, \tau) \Big( \partial_\tau - \mu + {\bm k^2 \over 2m} \Big) c_\sigma(\bm k, \tau) \nn
&&+ f_{\sigma}^{\dagger}(\tau) \Big( \partial_{\tau} - i \lambda(\tau) \Big) f_{\sigma}(\tau) + i N S \lambda(\tau) \nn
&&-{J_K\over N} c_\sigma^\dagger(\tau) f_\sigma(\tau) f_{\sigma'}^\dagger(\tau) c_{\sigma'}(\tau)  \bigg\} , \eqa
where the Fiertz identity for the inner product of Pauli spin matrices has been used and renormalization of the chemical potential at the impurity site has not been considered, not relevant for the Kondo effect. $\lambda(\tau)$ is a Lagrange multiplier variable to impose the above constraint, and $c_\sigma(\tau) = \int\!\frac{d^d\bm k}{(2\pi)^d} c_\sigma(\bm k, \tau)$ is an electron field at the impurity site.

Considering physical processes in the Kondo problem, it is natural to take a hybridization order parameter $b(\tau)$, where the partition function is given by
\begin{widetext}
\bqa Z &=& \int D c_{\sigma}(\bm{k},\tau) D f_{\sigma}(\tau) D b(\tau) D \lambda(\tau) \ e^{-S}, \nn S &=& \int_0^\beta\!d\tau \bigg\{ \int\!\frac{d^d\bm k}{(2\pi)^d} \ c_\sigma^\dagger(\bm k, \tau) \Big( \partial_\tau - \mu + {\bm k^2 \over 2m} \Big) c_\sigma(\bm k, \tau) + f_{\sigma}^{\dagger}(\tau) \Big( \partial_{\tau} - i \lambda(\tau) \Big) f_{\sigma}(\tau) + i N S \lambda(\tau) \nn
&& - b(\tau) f_\sigma^\dagger(\tau) c_\sigma(\tau)-  b^\dagger(\tau) c_\sigma^\dagger(\tau) f_\sigma(\tau) + {N\over J_K} b^\dagger(\tau) b(\tau) \bigg\} . \label{Effective_Theory_Kondo} \eqa
\end{widetext}
It is straightforward to see
\bqa && b(\tau) = \frac{J_{K}}{N} \Big\langle c_\sigma^\dagger(\tau) f_{\sigma}(\tau) \Big\rangle \eqa
in the mean-field approximation, which explains why this bosonic variable is called the hybridization order parameter.

Performing the functional integration for conduction electron fields, we obtain an effective quantum-mechanics theory for the Kondo effect
\begin{widetext}
\bqa Z &=& Z_{c} \int D f_{\sigma}(\tau) D b(\tau) D \lambda(\tau) \ \exp\bigg[ - \int_0^\beta\!d\tau \bigg\{ \int_0^\beta\! d\tau' \ f_\sigma^\dagger(\tau) b(\tau) G_c(\tau - \tau') b^\dagger(\tau') f_\sigma(\tau') \nn && + f_{\sigma}^{\dagger}(\tau) \Big(\partial_{\tau} - i \lambda(\tau) \Big) f_{\sigma}(\tau)
+ i N S \lambda(\tau) + \frac{N}{J_{K}} b^{\dagger}(\tau) b(\tau) \bigg\} \bigg] , \label{QM_Kondo} \eqa
\end{widetext}
where $Z_{c}$ is the partition function of conduction electrons and
\bqa && G_c(\tau) = {1\over\beta}\sum_{i\omega} e^{-i\omega\tau} \ G_c(i\omega), \nn && G_{c}(i\omega) = \int\!\frac{d^{d} \bm{k}}{(2\pi)^{d}} \frac{1}{i \omega + \mu - \frac{\bm{k}^{2}}{2m}} = - i \pi N_{F} \mbox{sign}(\omega) \nonumber \eqa
is the electron Green's function with the density of states $N_{F}$. If we perform the saddle-point approximation for the hybridization order parameter, we obtain a mean-field theory for the Kondo effect \cite{Kondo_Textbook}. Unfortunately, this mean-field theory gives rise to a continuous phase transition, regarded to be an artifact of the mean-field theory. In order to overcome this artifact, $1/N$ corrections have been introduced into the mean-field theory \cite{Kondo_1_N_Expansion}. It turns out that the hybridization order parameter vanishes due to $1/N$ corrections. On the other hand, the local Fermi-liquid physics has been claimed to be still preserved, demonstrated by thermodynamic properties. In this study we propose how to solve this strongly coupled quantum mechanics problem in a non-perturbative way, where the meaning of the non-perturbative way will be clarified in subsection \ref{Nonperturbative_Nature}.

\section{Main resutls}

\subsection{An effective Landau-Ginzburg field theory for the Kondo effect: Emergence of an extra dimension}
\label{subsection:main_action}

We find an effective field theory
\begin{widetext}
\bqa && Z = Z_{c} Z_{h}^{f} \int D f_{\sigma}(\tau) D b(\tau,z) \exp\bigg[ - \int_{0}^{\beta} d \tau \bigg\{ \int_{0}^{\beta} d \tau' f_{\sigma}^{\dagger}(\tau) b(\tau,z_{f}) G_{c}(\tau-\tau') b^{\dagger}(\tau',z_{f}) f_{\sigma}(\tau') + f_{\sigma}^{\dagger}(\tau) (\partial_{\tau} - \lambda) f_{\sigma}(\tau) \nn && + N S \lambda + \frac{N}{J_{K}} b^{\dagger}(\tau,0) b(\tau,0) \bigg\}
- \int_{0}^{z_{f}} d z \int_{0}^{\beta} d \tau \bigg\{ \frac{\pi N \Lambda_b
}{J_{K}} \Big( \partial_{z} b^{\dagger}(\tau,z) \Big) \Big( \partial_{z} b(\tau,z) \Big) + \frac{2 N S N_{F}}{v_{F} \Lambda_c^2} b(\tau,z) \partial_{\tau} b^{\dagger}(\tau,z) \bigg\} \bigg] . \label{Effective_Landau_Ginzburg_Theory} \eqa
\end{widetext}
Here, $Z_{h}$ is the partition function, which results from contributions of high-energy fluctuating fields in the Wilsonian renormalization group approach. $z$ is an extra dimension, where $z_{f} \rightarrow 1/\Lambda$ corresponds to the low-energy limit. In the Wilsonian renormalization group approach, high energy fluctuations in the momentum space of $\Lambda - d \Lambda < k < \Lambda$ are integrated out to renormalize the dynamics of low-energy excitations. It turns out that $d z$ is identified with $d \Lambda/\Lambda^2$. $v_{F}$ is the Fermi velocity and $\Lambda_{c}$ ($\Lambda_{b}$) is the high-momentum (high-frequency) cutoff of conduction electrons (hybridization order parameter) in the Wilsonian renormalization group analysis. All details are in section \ref{Derivation_Landau_Ginzburg}. We point out that the saddle-point approximation for the hybridization order parameter is taken into account, essentially the same as its Gaussian integration. As a result, we obtain an effective Landau-Ginzburg field theory for the hybridization order parameter after the Gaussian integration of the fermion variable, which appears with an extra dimension. Below, we claim that this effective field theory introduces not only $1/N$ but also quantum corrections of all orders in the $1/N$ expansion.

Considering the variation of an effective energy functional with respect to $b(\tau,z)$, we obtain an equation of motion for the hybridization order parameter field
\bqa && - \partial_{z}^{2} b^{\dagger}(\tau,z) + \frac{2 S N_{F} J_{K}}{\pi v_{F} \Lambda_b\Lambda_c^2} \partial_{\tau} b^{\dagger}(\tau,z) = 0 . \label{Diffusion_Equation} \eqa
In addition, we can also obtain the similar relation for $ b (\tau,z)$ with an additional minus sign, which corresponds to the Hermitian conjugation of Eq. (\ref{Diffusion_Equation}). The above equation represents the diffusion equation in one dimension. Appearance of the diffusion equation in this extra dimension has interesting physical implication, which will be discussed in section \ref{Solution_Kodno_Effect}.

The UV boundary condition is given by
\bqa && - \partial_{z} b^{\dagger}(\tau,z) \Big|_{z = 0} + {1 \over \pi \Lambda_b}b^{\dagger}(\tau,0)
= 0 , \label{UV_Boundary_Condition} \eqa
where the linear derivative in $z$ results from the boundary term of the second-order derivative in $z$. The IR boundary condition is
\bqa && \frac{\pi \Lambda_b}{J_{K}} \partial_{z} b^{\dagger}(\tau,z) \Big|_{z = z_{f}} \nn && + \int_{0}^{\beta} d \tau' G_{c}(\tau-\tau') b^{\dagger}(\tau',z_{f}) G_{f}(\tau' - \tau) = 0 , \label{IR_Boundary_Condition} \eqa
where the spinon Green's function $G_{f}(\tau - \tau') \equiv - \Big\langle T_{\tau} [ f_{\sigma}(\tau) f_{\sigma}^{\dagger}(\tau') ] \Big\rangle$ is given by the solution of
\bqa && (\partial_{\tau} - \lambda) G_{f}(\tau - \tau') \nn && + b(\tau,z_{f}) \int_{0}^{\beta} d \tau'' G_{c}(\tau-\tau'') b^{\dagger}(\tau'',z_{f}) G_{f}(\tau'' - \tau') \nn && = - \delta(\tau-\tau') . \label{Spinon_Green_Function} \eqa
$T_{\tau}$ is the time-ordering operator and the spin summation is not performed. The Lagrange multiplier is determined by the spinon number-constraint
\bqa && G_{f}(\tau \rightarrow 0) = S , \label{Number_Constraint} \eqa
where $i \lambda$ is replaced with $\lambda$ for the saddle-point analysis. We solve these coupled equations [Eqs. (\ref{Diffusion_Equation}), (\ref{UV_Boundary_Condition}), (\ref{IR_Boundary_Condition}), and (\ref{Number_Constraint}) with Eq. (\ref{Spinon_Green_Function})] in section \ref{Solution_Kodno_Effect} after presenting the derivation of our effective Landau-Ginzburg field theory with an extra dimension in section \ref{Derivation_Landau_Ginzburg}.

\subsection{The Kondo effect in the non-perturbative field theoretical framework: Impurity thermodynamics}
\label{Kondo_summary}

It is interesting to observe that the effective Landau-Ginzburg theory of Eq. (\ref{Effective_Landau_Ginzburg_Theory}) allows the saddle point approximation for the hybridization field $b(z,\tau)$ in the large$-N$ limit, giving rise to Eqs. (\ref{Diffusion_Equation}), (\ref{UV_Boundary_Condition}), and (\ref{IR_Boundary_Condition}). Surprisingly, this large$-N$ limit can describe the non-perturbative physics of the Kondo effect, as will be shown below. In other words, summing all loop quantum corrections organized in the $1/N$ expansion gives rise to completely a different large$-N$ effective Landau-Ginzburg theory, the large$-N$ limit of which describes a non-perturbative saddle point.

Solving the classical diffusion equation of Eq. (\ref{Diffusion_Equation}) with both boundary conditions of Eqs. (\ref{UV_Boundary_Condition}) and (\ref{IR_Boundary_Condition}), and substituting the solution into Eq. (\ref{Effective_Landau_Ginzburg_Theory}), we find an effective free energy functional for the Kondo effect, the most singular part of which is given by
\bqa && F_{imp} = - \frac{N}{\beta} \sum_{i \omega} \ln \bigg[ - i \omega - \lambda + C - i\pi N_F {\rm sign}(\omega) \nn && \times \Big\{ \alpha\sqrt{\frac{C}{D}} \cos \Big( \sqrt{\frac{C}{D}} z_{f} \Big) + \sin \Big( \sqrt{\frac{C}{D}} z_{f} \Big) \Big\}^{2} \bigg] \nn && + N S \lambda . \label{Impurity_Thermodynamics_Singular_Part} \eqa
Here, the Gaussian integration for the spinon variable in Eq. (\ref{Effective_Landau_Ginzburg_Theory}) has been performed. $D=\frac{\pi v_F \Lambda_b \Lambda_c^2}{2SN_F J_K}$ is an effective diffusion coefficient of Eq. (\ref{Diffusion_Equation}), and $C$ and $\alpha$ are constants of the hybridization order parameter field determined by self-consistent equations [Eq. (\ref{Self_Consistent_Equations})].
This expression implies that an effective Kondo temperature is
\bqa T_{K} = \pi N_{F} \Big\{ \alpha\sqrt{\frac{C}{D}} \cos \Big( \sqrt{\frac{C}{D}} \Big) + \sin \Big( \sqrt{\frac{C}{D}} \Big) \Big\}^{2}, \eqa
where $C$ is the solution of Eq. (\ref{eq:SCE_C}).

\begin{figure}[t!]
\includegraphics[width=7.5cm]{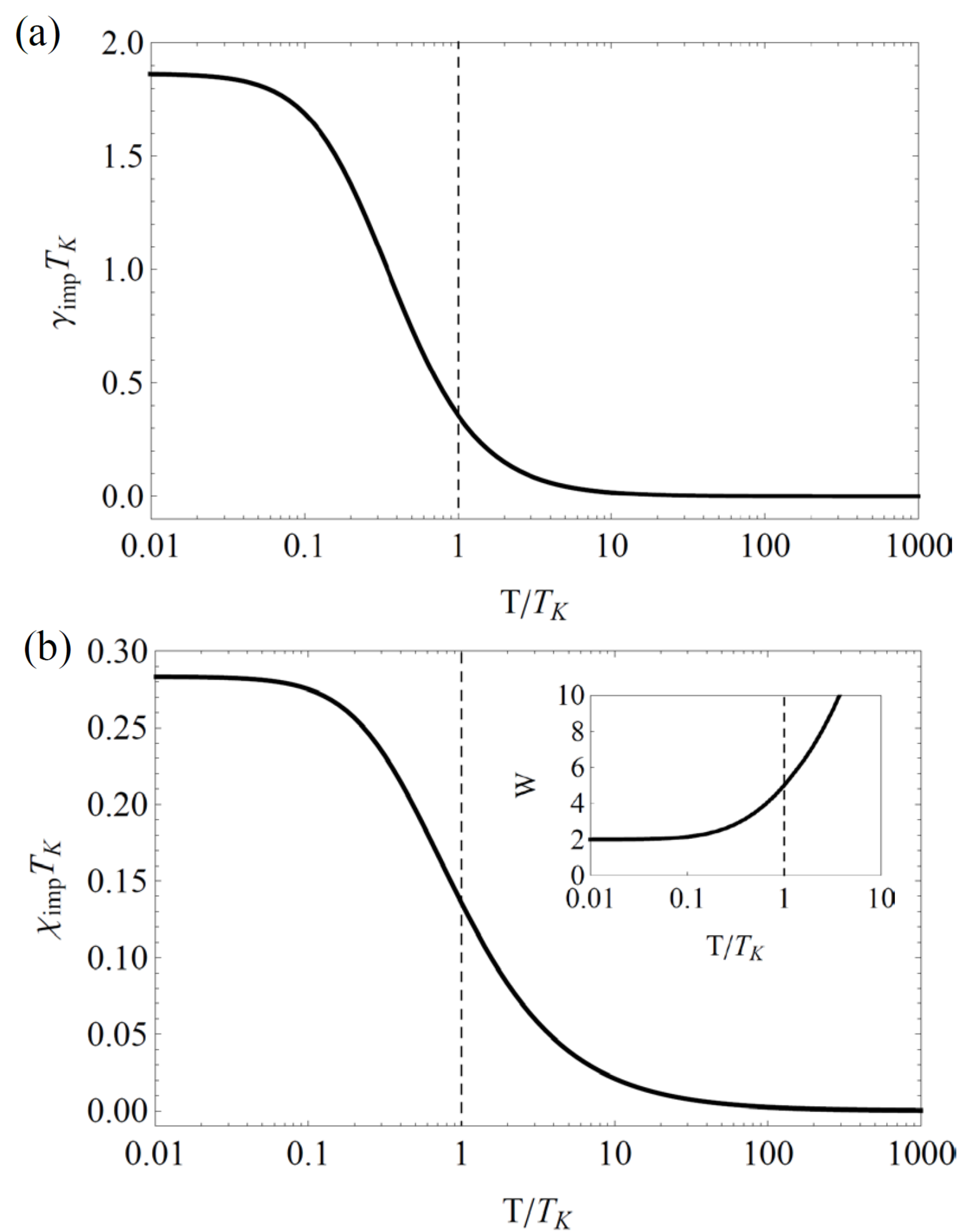}
\caption{ (a) Linear-log plot of the impurity specific heat coefficient [Eq. (\ref{eq:SH})] as a function of temperature. (b) Linear-log plot of the impurity spin susceptibility [Eq. (\ref{eq:SS})] as a function of temperature. The Wilson ratio is plotted in the inset. The vertical dotted line denotes $T/T_K = 1$.}
\label{Fig:TD}
\end{figure}

It is straightforward to obtain the specific heat coefficient $\gamma_{imp}(T)$ and the impurity spin susceptibility $\chi_{imp}(T)$, given by
\begin{eqnarray}
\gamma_{imp}(T) &\approx& \frac{N}{\pi T_K}\int_{-\infty}^{\infty} d \omega \frac{\partial^{2} f(\omega)}{\partial T^{2}} \tan^{-1} \Big( \frac{1}{\omega} \Big) \label{eq:SH},
\end{eqnarray}
and
\begin{eqnarray}
\chi_{imp}(T) &\approx& - \frac{N}{\pi T_K} \int_{-\infty}^{\infty} d \omega f(\omega) \frac{ \omega }{ ( \omega^{2} + 1)^{2} } \label{eq:SS}.
\end{eqnarray}
In the zero temperature limit, we have
\begin{eqnarray}
&& \gamma_{imp} \approx {\pi N\over 3} \frac{1}{T_{K}}, \ \ \chi_{imp} \approx \frac{N}{2\pi} \frac{1}{T_{K}} ,
\end{eqnarray}
typical for the Kondo effect \cite{Kondo_Textbook}. Indeed, we reproduce the local Fermi-liquid physics, given by the Wilson ratio at zero temperature
\bqa && W = {4\pi^2\over3}\frac{\chi_{imp}}{\gamma_{imp}} = 2 . \eqa

\begin{figure}[t!]
\includegraphics[width=7.5cm]{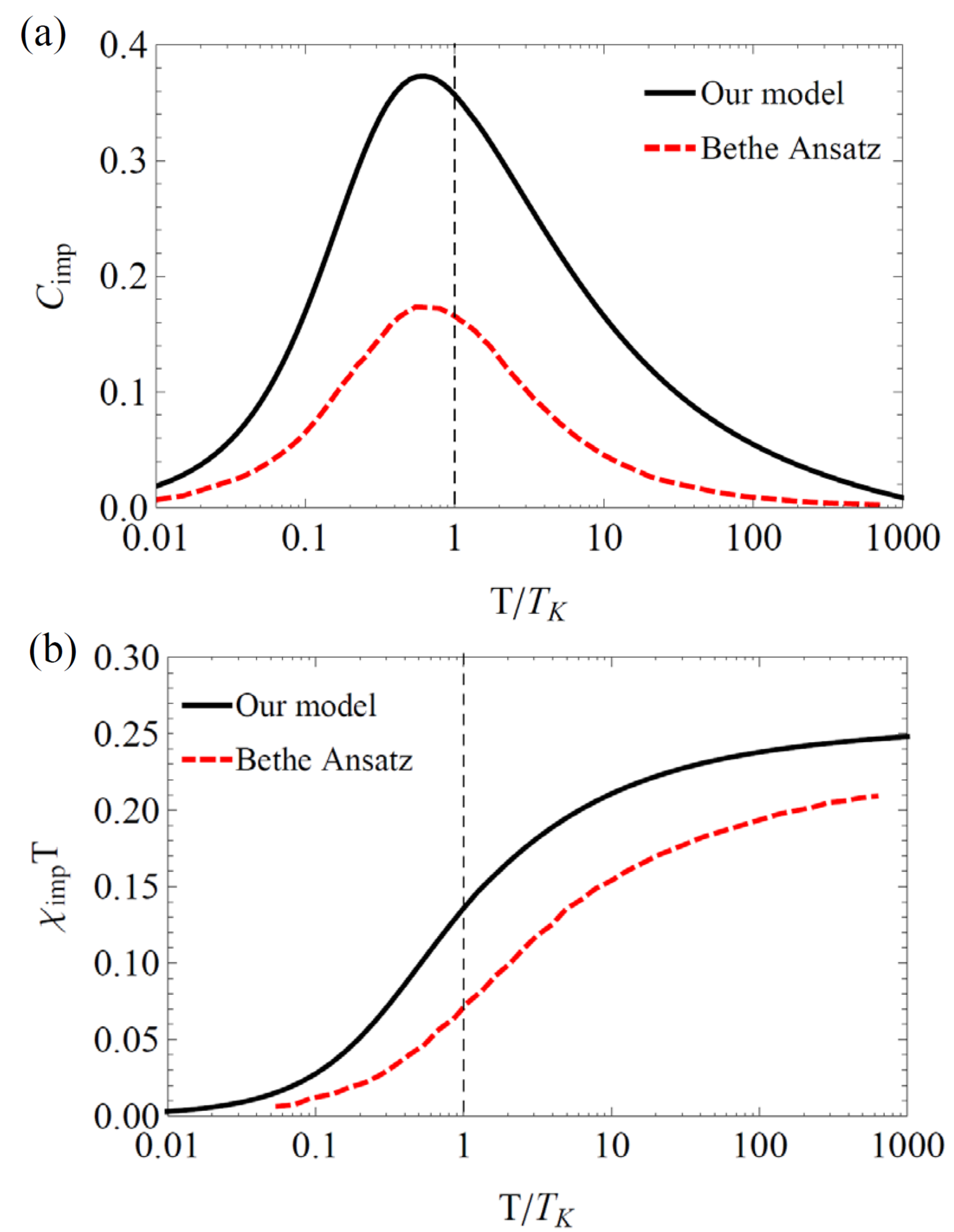}
\caption{ (a) Linear-log plot of the impurity specific heat as a function of temperature. (b) Linear-log plot of the impurity spin susceptibility multiplied by temperature $T$. Solid black (red dotted) lines are our (Bethe ansatz) results. }
\label{Fig:TD1}
\end{figure}

At finite temperatures, we solve Eq. (\ref{eq:SCE_C}) and perform the $\omega$-integration in Eqs. (\ref{eq:SH}) and (\ref{eq:SS}) numerically. Details are shown in Sec. \ref{Solution_Kodno_Effect}. Figure \ref{Fig:TD} shows that our effective field theory describes the crossover behavior from the decoupled local moment fixed point to the local Fermi-liquid fixed point quite successfully. In order to confirm this aspect, we compare our results with the Bethe ansatz solution for the single impurity Kondo model \cite{Kondo_Textbook,Bethe_Ansatz}. See Fig. \ref{Fig:TD1}. Although both the high and low temperature limits in the impurity thermodynamics coincide between our non-perturbative effective field theory and the Bethe ansatz, there exist discrepancies in the vicinity of the Kondo temperature, i.e., the crossover regime. We point out that the renormalization group transformation does not take into account the wave-function renormalization for the impurity fermion variable, as will be discussed below. We suspect that this poor men's scheme for renormalization is responsible for such discrepancies.

\section{Derivation of the non-perturbative field theory with an extra dimension for the Kondo effect: Continuous applications of Wilsonian renormalization group transformations} \label{Derivation_Landau_Ginzburg}

\subsection{The first iteration of Wilsonian renormalization group transformations}

We recall Eq. (\ref{Effective_Theory_Kondo})
\bqa Z &=& \int D c_{\sigma}(\bm{k},\tau) D f_{\sigma}(\tau) D \lambda(\tau) D b^{(0)}(\tau) \ e^{-S}, \nn S &=& \int_0^\beta\!d\tau \bigg\{\int \!\frac{d^{d} \bm{k}}{(2\pi)^{d}} \ c_{\sigma}^{\dagger}(\bm{k},\tau) \Big( \partial_{\tau} - \mu + \frac{\bm{k}^{2}}{2m} \Big) c_{\sigma}(\bm{k},\tau) \nn
&&+ f_{\sigma}^{\dagger}(\tau) \Big( \partial_{\tau} - i \lambda(\tau) \Big) f_{\sigma}(\tau) + i N S \lambda(\tau) \nn
&&- {b^{(0)}}(\tau) f_\sigma^\dagger(\tau) c_\sigma(\tau) - {b^{(0)}}^\dagger(\tau) c_\sigma^\dagger(\tau) f_\sigma(\tau) \nn
&& + {N\over J_K} {b^{(0)}}^\dagger(\tau) b^{(0)}(\tau) \bigg\} , \nonumber \eqa
where the superscript $(0)$ has been introduced into the hybridization order parameter.

An essential point of the present study is introducing quantum corrections into the mean-field theory through continuous applications of Wilsonian renormalization group transformations. However, it is not straightforward to deal with discrete spectrum at finite temperatures in the renormalization group analysis. In order to implement Wilsonian renormalization group transformations beyond the mean-field theory, we focus on the zero temperature limit, i.e., $\beta \rightarrow \infty$ as follows
\bqa W &=& \int D c_{\sigma}(\bm{k},\tau) D f_{\sigma}(\tau) D \lambda(\tau) D b^{(0)}(\tau) \ e^{-S}, \nn
S &=& \int_0^\infty\!d\tau \bigg\{ \int \!\frac{d^{d} \bm{k}}{(2\pi)^{d}} \ c_{\sigma}^{\dagger}(\bm{k},\tau) \Big( \partial_{\tau} - \mu + \frac{\bm{k}^{2}}{2m} \Big) c_{\sigma}(\bm{k},\tau) \nn
&&+ f_{\sigma}^{\dagger}(\tau) \Big( \partial_{\tau} - i \lambda(\tau) \Big) f_{\sigma}(\tau) + i N S \lambda(\tau) \nn
&&- b^{(0)}(\tau) f_\sigma^\dagger(\tau) c_\sigma(\tau) - {b^{(0)}}^\dagger(\tau) c_\sigma^\dagger(\tau) f_\sigma(\tau)  \nn
&& + {N\over J_K} {b^{(0)}}^\dagger(\tau) b^{(0)}(\tau) \bigg\} . \label{eq:zeroth} \eqa
Here, we still work in the Euclidean time.

To prepare for the Wilsonian renormalization group transformation, we separate all field variables into low- and high-energy degrees of freedom. For the conduction electrons, we set
\bqa c_{\sigma \omega}(\bm{k}) &=& c_{\sigma \omega}(\bm{k}) \Theta[(k - \Lambda_{c} + d \Lambda_{c}) (- \Lambda_{c} + d \Lambda_{c} - k)] \nn &+& c_{\sigma \omega}(\bm{k}) \Theta[(\Lambda_{c} - |k|) (|k| - \Lambda_{c} + d \Lambda_{c})] \nn &\equiv& c_{\sigma \omega}^{l}(\bm{k}) + c_{\sigma \omega}^{h}(\bm{k}) , \eqa
where $\Lambda_{c}$ is a momentum cutoff and $k$ is a momentum transverse to the Fermi surface. More precisely, we considered $\varepsilon_{\bm{k}} = \frac{\bm{k}^{2}}{2m} - \mu \approx \bm{v}_{F} \cdot (\bm{k} - \bm{k}_{F}) \equiv v_{F} k$ as usual, where $\bm{k}_{F}$ is a Fermi momentum and $\bm{v}_{F} = \bm{k}_{F} / m$ is a Fermi velocity. High$-$energy electron fields living within the momentum window of $\Lambda_{c} - d \Lambda_{c} < |k| < \Lambda_{c}$ are integrated over to renormalize hybridization fluctuations. Low$-$ and high$-$ energy degrees of freedom for spinon, holon (hybridization), and spinon chemical potential are given by
\bqa f_{\sigma \omega} &=& f_{\sigma \omega} \Theta(\Lambda_{f} - d \Lambda_{f} - |\omega|) \nn &+& f_{\sigma \omega} \Theta[(\Lambda_{f} - |\omega|)(|\omega| - \Lambda_{f} + d \Lambda_{f})] \nn &\equiv& f_{\sigma \omega}^{l} + f_{\sigma \omega}^{h} , \nonumber \eqa
\bqa b_{\omega} &=& b_{\omega} \Theta(\Lambda_{b} - d \Lambda_{b} - |\omega|) \nn &+& b_{\omega} \Theta[(\Lambda_{b} - |\omega|)(|\omega| - \Lambda_{b} + d \Lambda_{b})] \nn &\equiv& b_{\omega}^{l} + b_{\omega}^{h} , \nonumber \eqa
\bqa \lambda_{\omega-\omega'} &=& \lambda_{\omega-\omega'} \Theta(\Lambda_{\lambda} - d \Lambda_{\lambda} - |\omega-\omega'|) \nn &+& \lambda_{\omega-\omega'} \Theta[(\Lambda_{\lambda} - |\omega-\omega'|)(|\omega-\omega'| - \Lambda_{\lambda} + d \Lambda_{\lambda})] \nn &\equiv& \lambda_{\omega-\omega'}^{l} + \lambda_{\omega-\omega'}^{h} , \nonumber \eqa
essentially the same as the case of electron fields but in the frequency space. All integral regions given by $d \Lambda_{c}$, $d \Lambda_{f}$, $d \Lambda_{b}$, and $d \Lambda_{\lambda}$ are set to be equal, i.e., $v_{F} d \Lambda_{c} =  d \Lambda_{f} = d \Lambda_{b} = d \Lambda_{\lambda} \equiv d \Lambda$.

Integrating over high-energy variables and taking rescaling of all low-energy fields to return the cutoff into an original value, we obtain
\begin{widetext}
\bqa W &=& W_{h} \int D c_{\sigma}(\bm{k},\tau) D f_{\sigma}(\tau) D \lambda(\tau) D b^{(0)}(\tau) \ e^{-S}, \nn
S &=& \int_0^\infty\!d\tau \bigg\{ \int \!\frac{d^{d} \bm{k}}{(2\pi)^{d}} \ c_{\sigma}^{\dagger}(\bm{k},\tau) \Big( \partial_{\tau} - \mu + \frac{\bm{k}^{2}}{2m} \Big) c_{\sigma}(\bm{k},\tau) + f_{\sigma}^{\dagger}(\tau) \Big( \partial_{\tau} - i \lambda(\tau) \Big) f_{\sigma}(\tau) + i N S \lambda(\tau) \nn
&& - b^{(0)}(\tau) f_\sigma^\dagger(\tau) c_\sigma(\tau) - {b^{(0)}}^\dagger(\tau) c_\sigma^\dagger(\tau) f_\sigma(\tau) + {N\over J_K} {b^{(0)}}^\dagger(\tau) b^{(0)}(\tau) \nn
&& + g_{c} b^{(0)}(\tau) f_{\sigma}^{\dagger}(\tau) \partial_{\tau} \Big( f_{\sigma}(\tau) b^{(0) \dagger}(\tau) \Big) + g_{h}^{f} b^{(0)}(\tau) c_{\sigma}(\tau) c_{\sigma}^{\dagger}(\tau) b^{(0) \dagger}(\tau) - g_{h}^{b} f_{\sigma'}^{\dagger}(\tau) c_{\sigma'}(\tau) c_{\sigma}^{\dagger}(\tau) f_{\sigma}(\tau) \bigg\} . \label{After_Integration} \eqa
\end{widetext}
Compared to Eq. (\ref{eq:zeroth}), the integration over high-energy fields gives rise to the last three terms in the effective action with the multiplication of a factor $W_h$ into the partition function. Here, $W_{h}$ is the partition function of high$-$energy degrees of freedom. Coefficients are given by $g_c = 2{N_F\over v_F}{d\Lambda\over \Lambda_c^2} $, $g_h^f = i\lambda_{\omega=0} {d\Lambda \over \pi \Lambda_f^3}$, and $g_h^b = {J_K\over N}{d\Lambda\over\pi\Lambda_b}$. An essential aspect of this effective field theory is that locality in time is preserved in spite of the Wilsonian renormalization group transformation. Since this is an important point in the present study, we discuss this issue more carefully.

Consider the Kondo vertex with the high-energy mode of the $b$-field
\bqa && S_{int}^{K} = -\int \frac{d \omega}{2\pi} \int \frac{d \omega'}{2\pi} b_{\omega'}^{(0) h} f_{\sigma \omega}^{l\dagger} c_{\sigma \omega-\omega'}^{l} \nn && - \int \frac{d \omega}{2\pi} \int \frac{d \omega'}{2\pi} b_{\omega'}^{(0) h \dagger} c_{\sigma \omega-\omega'}^{l \dagger}  f_{\sigma \omega}^{l}  . \eqa
Performing the Gaussian integration for high-frequency holon (hybridization) fields in the second cumulant, we obtain
\begin{widetext}
\bqa \mathcal{S}_{K}^{(2)} &=& - \frac{1}{2} \Big( \Big\langle S_{int}^{K 2} \Big\rangle - \Big\langle S_{int}^{K} \Big\rangle^{2} \Big) = -\int_{\Lambda_b/b}^{\Lambda_b} \frac{d \omega'}{2\pi} \Big( \int \frac{d \omega}{2\pi} f_{\sigma \omega}^{l\dagger} c_{\sigma \omega-\omega'}^{l} \Big) g_{h}^{b}(i\omega') \Big( \int \frac{d \omega''}{2\pi} c_{\sigma' \omega''-\omega'}^{l \dagger}f_{\sigma' \omega''}^{l} \Big) \nn &=& - \frac{{J}_{K}}{N} {d \Lambda\over 2\pi} \Big( \int^{\Lambda_b/b} \frac{d \omega}{2\pi} f_{\sigma \omega}^{l\dagger} c_{\sigma \omega-\Lambda_{b}}^{l} \Big) \Big( \int^{\Lambda_b/b} \frac{d \omega''}{2\pi} c_{\sigma' \omega''-\Lambda_{b}}^{l \dagger} f_{\sigma' \omega''}^{l} \Big) + (\Lambda_b \rightarrow -\Lambda_b), \eqa
where $g_h^b(i\omega) = -\big< b_{\omega}^{(0)h} b_{\omega}^{(0)h\dagger} \big> = -{J_K\over N}$ is the high-frequency holon propagator and $b \approx 1+\delta\Lambda/\Lambda$. Rescaling the frequency $\omega \rightarrow \omega/b$, the cut off is restored to $\Lambda_b$, which results in
\bqa && \mathcal{S}_{K}^{(2)} = - \frac{{J}_{K}}{N} {d \Lambda\over 2\pi} \Big( \int^{\Lambda_b} \frac{d \omega}{2\pi} f_{\sigma \omega}^{l\dagger} c_{\sigma \omega-\Lambda_{b}}^{l} \Big) \Big( \int^{\Lambda_b} \frac{d \omega''}{2\pi} c_{\sigma' \omega''-\Lambda_{b}}^{l \dagger} f_{\sigma' \omega''}^{l} \Big) + (\Lambda_b \rightarrow -\Lambda_b) . \eqa
Here, we neglect higher order terms in $d\Lambda$. Fourier transforming to the time domain, we obtain
\bqa \mathcal{S}_{K}^{(2)} &=& - \frac{{J}_{K}}{N} {d \Lambda\over 2\pi} \int_{0}^{\infty} d \tau \int_{0}^{\infty} d \tau' e^{i \Lambda_{b} (\tau - \tau')} f_{\sigma}^{\dagger}(\tau) c_{\sigma}(\tau) c_{\sigma'}^{\dagger}(\tau') f_{\sigma'}(\tau') + (\Lambda_b \rightarrow -\Lambda_b) \nn &\approx& - \frac{{J}_{K}}{N} {d \Lambda\over\pi\Lambda_b} \int_{0}^{\infty} d \tau f_{\sigma}^{\dagger}(\tau) c_{\sigma}(\tau) c_{\sigma'}^{\dagger}(\tau) f_{\sigma'}(\tau). \eqa
An essential point is that the fast oscillating term given by $e^{\pm i \Lambda_{b} (\tau - \tau')}$ enforces the locality in the Kondo vertex.
\end{widetext}

One can show that the locality also holds for the other potential scattering term in the same way. $g_{h}^{f} b^{(0)}(\tau) c_{\sigma}(\tau) c_{\sigma}^{\dagger}(\tau) b^{(0) \dagger}(\tau)$ originates from the integration of high-frequency spinon variables in the Kondo-vertex term. This renormalization gives rise to non-magnetic potential scattering for low-energy electron fields, which has nothing to do with the Kondo effect.

The time derivative in $g_{c} b^{(0)}(\tau) f_{\sigma}^{\dagger}(\tau) \partial_{\tau} \big( f_{\sigma}(\tau) b^{(0) \dagger}(\tau) \big)$ comes from the propagator of high-energy conduction electrons. Consider the Kondo vertex with the high-energy mode of the $c$-field
\bqa && S_{int}^{imp} = -\int \frac{d \omega}{2\pi} \int \frac{d \omega'}{2\pi} b_{\omega'}^{(0) l} f_{\sigma \omega}^{l\dagger} \int \frac{d^{d} \bm{k}}{(2\pi)^{d}} c_{\sigma \omega-\omega'}^{h}(\bm{k}) \nn && - \int \frac{d \omega}{2\pi} \int \frac{d \omega'}{2\pi} b_{\omega'}^{(0) l \dagger} \int \frac{d^{d} \bm{k}}{(2\pi)^{d}} c_{\sigma \omega-\omega'}^{h \dagger}(\bm{k}) f_{\sigma \omega}^{l} . \eqa
Similar to the previous case, performing the integration for high$-$momentum conduction electron fields in the $d \Lambda \rightarrow 0$ limit and rescaling all low-frequency fields to recover the cutoff into an original value, we obtain
\begin{widetext}

\bqa \mathcal{S}_{imp}^{(2)} &=& - \frac{1}{2} \Big( \Big\langle S_{int}^{imp 2} \Big\rangle - \Big\langle S_{int}^{imp} \Big\rangle^{2} \Big) \nn &=& \int \frac{d \omega}{2\pi} \Big( \int \frac{d \omega'}{2\pi} b_{\omega'}^{(0) l} f_{\sigma \omega+\omega'}^{l\dagger} \Big) \int \frac{d^{d} \bm{k}}{(2\pi)^{d}} g_{h}(\bm{k},i \omega) \Big( \int \frac{d \omega''}{2\pi} f_{\sigma \omega+\omega''}^{l} b_{\omega''}^{(0) l \dagger} \Big) 
%
%
\nn  &\approx& 2 \frac{N_{F}}{v_{F}} \frac{d \Lambda}{\Lambda_{c}^{2}} \int \frac{d \omega}{2\pi} \Big( \int \frac{d \omega'}{2\pi} b_{\omega'}^{(0) l} f_{\sigma \omega+\omega'}^{l\dagger} \Big) (- i \omega) \Big( \int \frac{d \omega''}{2\pi} f_{\sigma \omega+\omega''}^{l} b_{\omega''}^{(0) l \dagger} \Big) \nn \overrightarrow{\mbox{Rescaling}} 
%
&& 2 \frac{N_{F}}{v_{F}} \frac{d \Lambda}{\Lambda_{c}^{2}} \int_{0}^{\infty} d \tau b^{(0)}(\tau) f_{\sigma}^{\dagger}(\tau) \partial_{\tau} \Big( f_{\sigma}(\tau) b^{(0) \dagger}(\tau) \Big) , \eqa
%
\end{widetext}
where \bqa && g_{h}(\bm{k},i \omega) \equiv - \langle c_{\sigma \omega}^{h}(\bm{k}) c_{\sigma \omega}^{h \dagger}(\bm{k}) \rangle = 2 N_{F} v_{F} d \Lambda \frac{- i \omega}{\omega^{2} + v_{F}^{2} \Lambda_{c}^{2}} \nonumber \eqa is the high$-$momentum electron propagator and $\omega \ll v_{F} \Lambda_{c}$ has been utilized from the second equality to the third \cite{Comment_on_Frequency_Domain}.

We note that the integration of the high$-$frequency spinon-chemical potential variable $\lambda_{\omega-\omega'}^{h}$ has not been introduced in this renormalization group procedure since this chemical potential renormalization for spinons does not give any serious effects on the Kondo effect. Of course, one can take into account its role in principle. Except for this aspect, our renormalization group procedure should be regarded as an ``exact" one up to the linear order of $d \Lambda$, where higher powers of $d \Lambda$ are all neglected. Actually, $d \Lambda$ is the control parameter in the integration procedure for high$-$energy degrees of freedom. The same strategy can be found in the functional renormalization group procedure \cite{Functional_RG}.

To finalize the first iteration of Wilsonian renormalization group transformations, we approximate $g_{c} b^{(0)}(\tau) f_{\sigma}^{\dagger}(\tau) \partial_{\tau} \big( f_{\sigma}(\tau) b^{(0) \dagger}(\tau) \big)$ in the following way
%
%
\bqa && g_{c} b^{(0)}(\tau) f_{\sigma}^{\dagger}(\tau) \partial_{\tau} \Big( f_{\sigma}(\tau) b^{(0) \dagger}(\tau) \Big) 
%
%
\nn &\approx& g_{c} b^{(0)}(\tau) f_{\sigma}^{\dagger}(\tau) f_{\sigma}(\tau) \partial_{\tau} b^{(0) \dagger}(\tau) . \eqa
%
%
$g_{c} b^{(0)}(\tau) b^{(0) \dagger}(\tau) f_{\sigma}^{\dagger}(\tau) \partial_{\tau} f_{\sigma}(\tau)$ is involved with wave-function renormalization for the spinon variable. Here, we do not take into account the wave-function renormalization for the spinon variable. In this respect this derivation is a poor man's version \cite{Kondo_Textbook}.

Inserting this time-derivative term for holon fields into the above partition function, Eq. (\ref{After_Integration}), and performing Hubbard-Stratonovich transformations with appropriate order parameters, we obtain
\begin{widetext}
\bqa
W &=& W_{h} \int D c_{\sigma}(\bm{k},\tau) D f_{\sigma}(\tau) D \lambda(\tau) D b^{(1)}(\tau) D \varphi^{(1)}(\tau) D n^{(1)}(\tau) D \psi^{(1)}(\tau) D \rho^{(1)}(\tau) D \delta b^{(1)}(\tau) \ e^{-S}, \nn
S &=& \int_{0}^{\infty}\! d \tau \bigg[ \int\!\frac{d^{d} \bm{k}}{(2\pi)^{d}} \  c_{\sigma}^{\dagger}(\bm{k},\tau) \Big( \partial_{\tau} - \mu + \frac{\bm{k}^{2}}{2m} \Big) c_{\sigma}(\bm{k},\tau) + f_\sigma^\dagger(\tau) \Big( \partial_\tau - i\lambda(\tau) \Big) f_\sigma(\tau) + iNS\lambda (\tau) \nn
&& - b^{(1)}(\tau) f_{\sigma}^{\dagger}(\tau) c_{\sigma}(\tau) - b^{(1) \dagger}(\tau) c_{\sigma}^{\dagger}(\tau) f_{\sigma}(\tau) + {N\over J_K} \Big( b^{(1)\dagger}(\tau) - \delta b^{(1)\dagger}(\tau) \Big) \Big( b^{(1)}(\tau) - \delta b^{(1)}(\tau) \Big) \nn
&& + g_c \Big( b^{(1)}(\tau) - \delta b^{(1)}(\tau) \Big) n^{(1)}(\tau) \partial_\tau \Big( b^{(1)\dagger}(\tau) - \delta b^{(1)\dagger}(\tau) \Big) + i\varphi^{(1)}(\tau) \Big( n^{(1)}(\tau) - f_\sigma^\dagger(\tau) f_\sigma(\tau) \Big) \nn
&& + g_h^f \Big( b^{(1)}(\tau) - \delta b^{(1)}(\tau) \Big) \rho^{(1)}(\tau) \Big( b^{(1)\dagger}(\tau) - \delta b^{(1)\dagger}(\tau) \Big) + i\psi^{(1)}(\tau) \Big( \rho^{(1)}(\tau) - c_\sigma(\tau) c_\sigma^\dagger(\tau) \Big) \nn
&& + {1\over g_h^b} \delta b^{(1)\dagger}(\tau) \delta b^{(1)}(\tau) \bigg] . \eqa
\end{widetext}
Here, $n^{(1)}(\tau) = f_{\sigma}^{\dagger}(\tau) f_{\sigma}(\tau)$ is introduced into the above expression, where the canonical conjugate variable $\varphi^{(1)}(\tau)$ plays the role of a Lagrange multiplier to impose this constraint. In the same way $\rho^{(1)}(\tau) = c_\sigma(\tau) c_\sigma^\dagger(\tau)$ is taken into account with its canonical conjugate pair $\psi^{(1)}(\tau)$. Another double-trace operator, $- g_{h}^{b} f_{\sigma'}^{\dagger}(\tau) c_{\sigma'}(\tau) c_{\sigma}^{\dagger}(\tau) f_{\sigma}(\tau)$, is decomposed into $-\delta b^{(1)}(\tau) f_{\sigma'}^{\dagger}(\tau) c_{\sigma'}(\tau) - \delta b^{(1) \dagger}(\tau) c_{\sigma}^{\dagger}(\tau) f_{\sigma}(\tau) + \frac{1}{g_{h}^{b}} \delta b^{(1) \dagger}(\tau) \delta b^{(1)}(\tau)$. In the above expression we defined $b^{(1)}(\tau) = b^{(0)}(\tau) + \delta b^{(1)}(\tau)$ and changed the integration field from $b^{(0)}(\tau)$ to $b^{(1)}(\tau)$.

\subsection{Continuous applications of Wilsonian renormalization group transformations}

One may repeat the previous renormalization group procedure: Separating low- and high-energy degrees of freedom for electron fields and others with the superscript index $(1)$, integrating over high-energy fields, and taking rescaling of all low-energy fields to return the cutoff into an original value, he/she will find the following expression of the partition function
\begin{widetext}
\bqa W &=& W_{h}^2 \int D c_{\sigma}(\bm{k},\tau) D f_{\sigma}(\tau) D \lambda(\tau) D b^{(1)}(\tau) D \varphi^{(1)}(\tau) D n^{(1)}(\tau) D \psi^{(1)}(\tau) D \rho^{(1)}(\tau) D \delta b^{(1)}(\tau) \ e^{-S}, \nn
S &=& \int_{0}^{\infty}\! d \tau \bigg[ \int\!\frac{d^{d} \bm{k}}{(2\pi)^{d}} \  c_{\sigma}^{\dagger}(\bm{k},\tau) \Big( \partial_{\tau} - \mu + \frac{\bm{k}^{2}}{2m} \Big) c_{\sigma}(\bm{k},\tau) + f_\sigma^\dagger(\tau) \Big( \partial_\tau - i\lambda(\tau) \Big) f_\sigma(\tau) + iNS\lambda (\tau) \nn
&& - b^{(1)}(\tau) f_{\sigma}^{\dagger}(\tau) c_{\sigma}(\tau) - b^{(1) \dagger}(\tau) c_{\sigma}^{\dagger}(\tau) f_{\sigma}(\tau) + {N\over J_K} \Big( b^{(1)\dagger}(\tau) - \delta b^{(1)\dagger}(\tau) \Big) \Big( b^{(1)}(\tau) - \delta b^{(1)}(\tau) \Big) \nn
&& + g_c \Big( b^{(1)}(\tau) - \delta b^{(1)}(\tau) \Big) n^{(1)}(\tau) \partial_\tau \Big( b^{(1)\dagger}(\tau) - \delta b^{(1)\dagger}(\tau) \Big) + i\varphi^{(1)}(\tau) \Big( n^{(1)}(\tau) - f_\sigma^\dagger(\tau) f_\sigma(\tau) \Big) \nn
&& - g_h^f \Big( b^{(1)}(\tau) - \delta b^{(1)}(\tau) \Big) \rho^{(1)}(\tau) \Big( b^{(1)\dagger}(\tau) - \delta b^{(1)\dagger}(\tau) \Big) + i\psi^{(1)}(\tau) \Big( \rho^{(1)}(\tau) - c_\sigma(\tau) c_\sigma^\dagger(\tau) \Big) \nn
&& + {1\over g_h^b} \delta b^{(1)\dagger}(\tau) \delta b^{(1)}(\tau) \nn
&& + g_{c} b^{(1)}(\tau) f_{\sigma}^{\dagger}(\tau) \partial_{\tau} \Big( f_{\sigma}(\tau) b^{(1) \dagger}(\tau) \Big) + g_{h}^{f} b^{(1)}(\tau) c_{\sigma}(\tau) c_{\sigma}^{\dagger}(\tau) b^{(1) \dagger}(\tau) - g_{h}^{b} f_{\sigma'}^{\dagger}(\tau) c_{\sigma'}(\tau) c_{\sigma}^{\dagger}(\tau) f_{\sigma}(\tau) \bigg]. \eqa
\end{widetext}
Note that, as before, we have three terms in the last line with the same coefficients, $g_c$, $g_h^f$, and $g_h^b$, but with the replacement of the superscript from $(0)$ to $(1)$ of the fields.

Neglecting the wave-function renormalization for the spinon variable and implementing Hubbard-Stratonovich transformations with appropriate order parameters once again, we find
\begin{widetext}
\bqa W &=& W_{h}^2 \int D c_{\sigma}(\bm{k},\tau) D f_{\sigma}(\tau) D \lambda(\tau) D b^{(2)}(\tau) \prod_{w=1}^2 \Big[ D \varphi^{(w)}(\tau) D n^{(w)}(\tau) D \psi^{(w)}(\tau) D \rho^{(w)}(\tau) D \delta b^{(w)}(\tau) \Big] \ e^{-S}, \nn
S &=& \int_{0}^{\infty}\! d \tau \bigg[ \int\!\frac{d^{d} \bm{k}}{(2\pi)^{d}} \  c_{\sigma}^{\dagger}(\bm{k},\tau) \Big( \partial_{\tau} - \mu + \frac{\bm{k}^{2}}{2m} \Big) c_{\sigma}(\bm{k},\tau) + f_\sigma^\dagger(\tau) \Big( \partial_\tau - i\lambda(\tau) \Big) f_\sigma(\tau) + iNS\lambda (\tau) \nn
&& -b^{(2)}(\tau) f_{\sigma}^{\dagger}(\tau) c_{\sigma}(\tau) - b^{(2) \dagger}(\tau) c_{\sigma}^{\dagger}(\tau) f_{\sigma}(\tau)
 + {N\over J_K} \Big( b^{(2)\dagger}(\tau) - \delta b^{(2)\dagger}(\tau) - \delta b^{(1)\dagger}(\tau) \Big) \Big( b^{(2)}(\tau) - \delta b^{(2)}(\tau) - \delta b^{(1)}(\tau) \Big) \nn
&& + g_c \Big( b^{(2)}(\tau) - \delta b^{(2)}(\tau) - \delta b^{(1)}(\tau) \Big) n^{(1)}(\tau) \partial_\tau \Big( b^{(2)\dagger}(\tau) - \delta b^{(2)\dagger}(\tau) - \delta b^{(1)\dagger}(\tau) \Big) + i\varphi^{(1)}(\tau) \Big( n^{(1)}(\tau) - f_\sigma^\dagger(\tau) f_\sigma(\tau) \Big) \nn
&& + g_h^f \Big( b^{(2)}(\tau) - \delta b^{(2)}(\tau) - \delta b^{(1)}(\tau) \Big) \rho^{(1)}(\tau) \Big( b^{(2)\dagger}(\tau) - \delta b^{(2)\dagger}(\tau) - \delta b^{(1)\dagger}(\tau) \Big) + i\psi^{(1)}(\tau) \Big( \rho^{(1)}(\tau) - c_\sigma(\tau) c_\sigma^\dagger(\tau) \Big) \nn
&& + {1\over g_h^b} \delta b^{(1)\dagger}(\tau) \delta b^{(1)}(\tau) \nn
&& + g_c \Big( b^{(2)}(\tau) - \delta b^{(2)}(\tau) \Big) n^{(2)}(\tau) \partial_\tau \Big( b^{(2)\dagger}(\tau) - \delta b^{(2)\dagger}(\tau) \Big) + i\varphi^{(2)}(\tau) \Big( n^{(2)}(\tau) - f_\sigma^\dagger(\tau) f_\sigma(\tau) \Big) \nn
&& + g_h^f \Big( b^{(2)}(\tau) - \delta b^{(2)}(\tau) \Big) \rho^{(2)}(\tau) \Big( b^{(2)\dagger}(\tau) - \delta b^{(2)\dagger}(\tau) \Big) + i\psi^{(2)}(\tau) \Big( \rho^{(2)}(\tau) - c_\sigma(\tau) c_\sigma^\dagger(\tau) \Big) \nn
&& + {1\over g_h^b} \delta b^{(2)\dagger}(\tau) \delta b^{(2)}(\tau) \bigg] , \eqa
where the integration field is changed from $b^{(1)}(\tau)$ to $b^{(2)}(\tau) = b^{(1)}(\tau) + \delta b^{(2)}(\tau)$.

Now, it is straightforward to implement Wilsonian renormalization group transformations continuously. As a result, we obtain
\bqa W &=& W_{h}^f \int D c_{\sigma}(\bm{k},\tau) D f_{\sigma}(\tau) D \lambda(\tau) D b^{(f)}(\tau) \prod_{w=1}^f \Big[ D \varphi^{(w)}(\tau) D n^{(w)}(\tau) D \psi^{(w)}(\tau) D \rho^{(w)}(\tau) D \delta b^{(w)}(\tau) \Big] \ e^{-S}, \nn
S &=& \int_{0}^{\infty}\!d\tau \bigg[ \int \frac{d^{d} \bm{k}}{(2\pi)^{d}} \  c_{\sigma}^{\dagger}(\bm{k},\tau) \Big( \partial_{\tau} - \mu + \frac{\bm{k}^{2}}{2m} \Big) c_{\sigma}(\bm{k},\tau) + f_\sigma^\dagger(\tau) \Big( \partial_\tau - i\lambda(\tau) \Big) f_\sigma(\tau) + iNS\lambda (\tau) \nn
&& -b^{(f)}(\tau) f_{\sigma}^{\dagger}(\tau) c_{\sigma}(\tau) - b^{(f) \dagger}(\tau) c_{\sigma}^{\dagger}(\tau) f_{\sigma}(\tau)
+ {N\over J_K} \bigg( b^{(f)\dagger}(\tau) - \sum_{w=1}^f \delta b^{(w)\dagger}(\tau) \bigg) \bigg( b^{(f)}(\tau) - \sum_{w=1}^f \delta b^{(w)}(\tau) \bigg) \nn
&& + g_c \sum_{w=1}^f \bigg( b^{(f)}(\tau) - \sum_{w'=w}^f b^{(w')}(\tau) \bigg) n^{(w)}(\tau) \partial_\tau \bigg( b^{(f)\dagger}(\tau) - \sum_{w'=w}^f b^{(w') \dagger} (\tau) \bigg) + \sum_{w=1}^f i\varphi^{(w)}(\tau) \Big( n^{(w)}(\tau) - f_\sigma^\dagger(\tau) f_\sigma(\tau) \Big) \nn
&&+g_h^f \sum_{w=1}^f \bigg( b^{(f)}(\tau) - \sum_{w'=w}^f b^{(w')}(\tau) \bigg) \rho^{(w)}(\tau) \bigg( b^{(f)\dagger}(\tau) - \sum_{w'=w}^f b^{(w') \dagger} (\tau) \bigg) + \sum_{w=1}^f i\psi^{(w)}(\tau) \Big( \rho^{(w)}(\tau) - c_\sigma(\tau) c_\sigma^\dagger(\tau) \Big) \nn
&& +{1\over g_h^b} \sum_{w=1}^f \delta b^{(w)\dagger}(\tau) \delta b^{(w)}(\tau) \bigg] . \eqa
\end{widetext}

The final step is changing the integration fields from $b^{(f)}(\tau)$ and $\delta b^{(w)}(\tau)$ with $w=1, \cdots, f$ to $b^{(0)}(\tau)$ and $b^{(w)}(\tau) = b^{(0)}(\tau) + \sum_{n=1}^w \delta b^{(n)}(\tau)$ with $w=1, \cdots, f$. As a result, the partition function and the effective action become
\begin{widetext}
\bqa W &=& W_{h}^f \int D c_{\sigma}(\bm{k},\tau) D f_{\sigma}(\tau) D \lambda(\tau) D b^{(0)}(\tau) \prod_{w=1}^f \Big[ D \varphi^{(w)}(\tau) D n^{(w)}(\tau) D \psi^{(w)}(\tau) D \rho^{(w)}(\tau) D b^{(w)}(\tau) \Big] \ e^{-S}, \nn
S &=& \int_{0}^{\infty}\!d\tau \bigg[ \int \frac{d^{d} \bm{k}}{(2\pi)^{d}} \  c_{\sigma}^{\dagger}(\bm{k},\tau) \Big( \partial_{\tau} - \mu + \frac{\bm{k}^{2}}{2m} \Big) c_{\sigma}(\bm{k},\tau) + f_\sigma^\dagger(\tau) \Big( \partial_\tau - i\lambda(\tau) \Big) f_\sigma(\tau) + iNS\lambda (\tau) \nn
&& -b^{(f)}(\tau) f_{\sigma}^{\dagger}(\tau) c_{\sigma}(\tau) - b^{(f) \dagger}(\tau) c_{\sigma}^{\dagger}(\tau) f_{\sigma}(\tau) +{N\over J_K} b^{(0)\dagger}(\tau) b^{(0)}(\tau) \nn
&& + g_c \sum_{w=1}^f b^{(w-1)}(\tau) n^{(w)}(\tau) \partial_\tau b^{(w-1)\dagger}(\tau) + \sum_{w=1}^f i\varphi^{(w)}(\tau) \Big( n^{(w)}(\tau) - f_\sigma^\dagger(\tau) f_\sigma(\tau) \Big) \nn
&&+g_h^f \sum_{w=1}^f b^{(w-1)}(\tau) \rho^{(w)}(\tau) b^{(w-1)\dagger}(\tau) + \sum_{w=1}^f i\psi^{(w)}(\tau) \Big( \rho^{(w)}(\tau) - c_\sigma(\tau) c_\sigma^\dagger(\tau) \Big) \nn
&& +{1\over g_h^b} \sum_{w=1}^f \Big( b^{(w)\dagger}(\tau) - b^{(w-1)\dagger}(\tau) \Big) \Big( b^{(w)}(\tau) - b^{(w-1)}(\tau) \Big) \bigg] . \label{eq:discrete} \eqa
\end{widetext}

\subsection{Emergence of an extra dimension} \label{dz_Equal_dLambda}

In order to confirm the emergence of an extra dimension, we reformulate the previous expression of the partition function as follows. First, we consider
\bqa && \sum_{w = 1}^{f} i \varphi^{(w)}(\tau) \Big( n^{(w)}(\tau) - f_\sigma^\dagger(\tau) f_\sigma(\tau) \Big) \nn
&& = \sum_{w = 1}^{f} a_{uv} {i \varphi^{(w)}(\tau) \over a_{uv}} \Big( n^{(w)}(\tau) - f_\sigma^\dagger(\tau) f_\sigma(\tau) \Big) \nn && = \int_{0}^{z_{f}}\! d z \ i \varphi(\tau,z) \Big( n(\tau,z) - f_\sigma^\dagger(\tau) f_\sigma(\tau) \Big) , \label{Sum_to_Integral} \eqa
where $a_{uv}$ is a length scale of UV cutoff, regarded to be lattice spacing in the $z$-direction. In the last line of the equation, we took into account $n(\tau, z=wa_{uv}) = n^{(w)}(\tau)$ and $\varphi(\tau, z=wa_{uv}) = \varphi^{(w)}(\tau)/a_{uv}$. The upper limit of the integration is $z_f = fa_{uv}$. Second, we consider
\bqa && \frac{1}{g_{h}^{b}} \sum_{w = 1}^f \Big( b^{(w) \dagger}(\tau) - b^{(w-1) \dagger}(\tau) \Big) \Big( b^{(w)}(\tau) - b^{(w-1)}(\tau) \Big) \nn && = {a_{uv} \over g_h^b} \sum_{w = 1}^f a_{uv} \frac{b^{(w) \dagger}(\tau) - b^{(w-1) \dagger}(\tau)}{a_{uv}} \frac{b^{(w)}(\tau) - b^{(w-1)}(\tau)}{a_{uv}}  \nn
&& = {1\over \tilde g_h^b} \int_0^{z_f}\!dz \ \left( \frac{\partial b^\dagger(\tau, z)}{\partial z} \right) \left( \frac{\partial b(\tau, z)}{\partial z} \right) , \label{Sum_to_Integral_dLambda}
\eqa
where $\tilde g_{h}^{b} = g_h^b/a_{uv}$ and $b(\tau, z=wa_{uv}) = b^{(w)}(\tau)$ have been introduced. In Eq. (\ref{Sum_to_Integral}) the cutoff length scale $a_{uv}$ seems to be determined arbitrarily. This is not true. We recall $g_h^b = {J_K\over N}{d\Lambda\over\pi\Lambda_b}$. Then, we have $1 / \tilde g_{h}^{b} \propto a_{uv} / d \Lambda$. We also point out that all these calculations are controlled in the $d \Lambda \rightarrow 0$ limit. In order to define Eq. (\ref{Sum_to_Integral_dLambda}) consistently in this controllable limit, we are forced to take the equation of $a_{uv} = c_{uv} d \Lambda$, where $c_{uv}$ is a positive constant, set to be $c_{uv} = 1$ for simplicity. Identifying $d \Lambda$ with $d z$, i.e., $d\Lambda = dz = a_{uv}$, we have $\tilde g_h^b = {J_K \over \pi N \Lambda_b}$. Equivalence between $d \Lambda$ and $d z$ indicates that the extra dimension may be regarded as the scale of renormalization group transformations. As a result, we reach the following expression for the partition function of the Kondo effect
\begin{widetext}
\bqa W &=& W_{h}^f \int D c_{\sigma}(\bm{k},\tau) D f_{\sigma}(\tau) D \lambda(\tau) D \varphi(\tau,z) D n(\tau,z) D \psi(\tau,z) D \rho(\tau,z) D b(\tau,z) \ e^{-S}, \nn
S &=& \int_{0}^{\infty}\!d\tau \bigg[ \int \frac{d^{d} \bm{k}}{(2\pi)^{d}} \  c_{\sigma}^{\dagger}(\bm{k},\tau) \Big( \partial_{\tau} - \mu + \frac{\bm{k}^{2}}{2m} \Big) c_{\sigma}(\bm{k},\tau) + f_\sigma^\dagger(\tau) \Big( \partial_\tau - i\lambda(\tau) \Big) f_\sigma(\tau) + iNS\lambda (\tau) \nn
&& - b(\tau,z_f) f_{\sigma}^{\dagger}(\tau) c_{\sigma}(\tau) - b^{\dagger}(\tau,z_f) c_{\sigma}^{\dagger}(\tau) f_{\sigma}(\tau) + {N\over J_K} b^{\dagger}(\tau,0) b(\tau,0) \nn
&& + \tilde g_c \int_0^{z_f}\!dz \ b(\tau,z) n(\tau,z) \partial_\tau b^{\dagger}(\tau,z) + \int_0^{z_f}\!dz \ i\varphi(\tau,z) \Big( n(\tau,z) - f_\sigma^\dagger(\tau) f_\sigma(\tau) \Big) \nn
&& +\tilde g_h^f \int_0^{z_f}\!dz \ b(\tau,z) \rho(\tau,z) b^{\dagger}(\tau,z) + \int_0^{z_f}\!dz \ i\psi(\tau,z) \Big( \rho(\tau,z) - c_\sigma(\tau) c_\sigma^\dagger(\tau) \Big) \nn
&& + {1\over \tilde g_h^b} \int_0^{z_f}\!dz \ \partial_z b^{\dagger}(\tau,z) \partial_z b(\tau,z) \bigg]. \eqa
\end{widetext}
Here, other fields and couplings are taken into account similarly as $\rho(\tau,z=wa_{uv}) = \rho^{(w)}(\tau)$, $\psi(\tau, z=wa_{uv}) = \psi^{(w)}(\tau)/a_{uv}$, $\tilde g_c = g_c/a_{uv} = 2N_F/(v_F\Lambda_c^2)$, and $\tilde g_h^f = g_h^f/a_{uv} = i\lambda_{w=0}/(\pi\Lambda_f^3)$.

Focusing on the Kondo effect, where non-magnetic potential scattering for electrons is neglected, we can simplify the above as
\begin{widetext}
\bqa W &=& W_{h}^f \int D c_{\sigma}(\bm{k},\tau) D f_{\sigma}(\tau) D \lambda(\tau) D \varphi(\tau,z) D n(\tau,z) D b(\tau,z) \ e^{-S}, \nn
S &=& \int_{0}^{\infty}\!d\tau \bigg[ \int \frac{d^{d} \bm{k}}{(2\pi)^{d}} \  c_{\sigma}^{\dagger}(\bm{k},\tau) \Big( \partial_{\tau} - \mu + \frac{\bm{k}^{2}}{2m} \Big) c_{\sigma}(\bm{k},\tau) + f_\sigma^\dagger(\tau) \Big( \partial_\tau - i\lambda(\tau) \Big) f_\sigma(\tau) + iNS\lambda (\tau) \nn
&& - b(\tau,z_f) f_{\sigma}^{\dagger}(\tau) c_{\sigma}(\tau) - b^{\dagger}(\tau,z_f) c_{\sigma}^{\dagger}(\tau) f_{\sigma}(\tau) + {N\over J_K} b^{\dagger}(\tau,0) b(\tau,0) \nn
&& + \tilde g_c \int_0^{z_f}\!dz \ b(\tau,z) n(\tau,z) \partial_\tau b^{\dagger}(\tau,z) + \int_0^{z_f}\!dz \ i\varphi(\tau,z) \Big( n(\tau,z) - f_\sigma^\dagger(\tau) f_\sigma(\tau) \Big) \nn
&& +{1\over \tilde g_h^b} \int_0^{z_f}\!dz \ \partial_z b^{\dagger}(\tau,z) \partial_z b(\tau,z) \bigg]. \eqa
\end{widetext}

Working out the path integral of $\int D \varphi(\tau,z)$, we obtain $n(\tau,z) = f_\sigma^\dagger(\tau) f_\sigma(\tau) = NS$ with the introduction of the number constraint. Taking the path integral of $\int D c_{\sigma}(\bm{k},\tau)$, we obtain the final expression of the partition function for the Kondo effect
\begin{widetext}
\bqa W &=& W_{h}^f Z_c \int D f_{\sigma}(\tau) D \lambda(\tau) D b(\tau,z) \ e^{-S}, \nn
S &=& \int_{0}^{\infty}\!d\tau\int_{0}^{\infty}\!d\tau' \ f_\sigma^\dagger(\tau) b(\tau, z_f) G_c(\tau-\tau') b^\dagger(\tau', z_f) f_\sigma(\tau') + \int_0^\infty\!d\tau \ \bigg[ f_\sigma^\dagger(\tau) \Big( \partial_\tau - i\lambda(\tau) \Big) f_\sigma(\tau) + iNS\lambda (\tau) \nn && + {N\over J_K} b^{\dagger}(\tau,0) b(\tau,0) \bigg] + \int_0^\infty\!d\tau \int_0^{z_f}\!dz \ \bigg[ NS\tilde g_c b(\tau,z) \partial_\tau b^{\dagger}(\tau,z) + {1\over \tilde g_h^b} \partial_z b^{\dagger}(\tau,z) \partial_z b(\tau,z) \bigg], \label{eq:final_action}\eqa
\end{widetext}
where the electron propagator is recalled as
\bqa && G_{c}(i\omega) \equiv - \int\!\frac{d^{d} \bm{k}}{(2\pi)^{d}} \ \Big\langle c_{\sigma}(\bm{k},i \omega) c_{\sigma}^{\dagger}(\bm{k},i \omega) \Big\rangle \nn && = \int\!\frac{d^{d} \bm{k}}{(2\pi)^{d}} \frac{1}{i \omega + \mu - \frac{\bm{k}^{2}}{2m}} = - i \pi N_{F} \mbox{sign}(\omega) \nonumber \eqa
in the frequency space. The spin summation is not performed. This effective action, Eq. (\ref{eq:final_action}), is one of the main results in this study, shown in Sec. \ref{subsection:main_action}.

\subsection{The role of the extra dimension in the effective Landau-Ginzburg field theory: Introduction of quantum corrections of all orders in the $1/N$ expansion} \label{Nonperturbative_Nature}

Based on the demonstration of this subsection, we claim that the effective Landau-Ginzburg field theory Eq. (\ref{Effective_Landau_Ginzburg_Theory}) takes into account quantum corrections of all orders in the scheme of the $1/N$ expansion.

\subsubsection{Discretization}

In order to figure out the role of the emergent extra dimension in the effective Landau-Ginzburg field theory, we discretize the $z-$coordinate as follows
\begin{widetext}
\bqa && Z = Z_{c} Z_{h}^{f} \int D f_{\sigma}(\tau) D b_0(\tau) \Pi_{w = 1}^{f} D b_{w}(\tau) \exp \bigg[ - \int_{0}^{\beta} d \tau \Big\{ \int_{0}^{\beta} d \tau' f_{\sigma}^{\dagger}(\tau) b_{f}(\tau) G_{c}(\tau-\tau') b_{f}^{\dagger}(\tau') f_{\sigma}(\tau') \nn && + f_{\sigma}^{\dagger}(\tau) (\partial_{\tau} - \lambda) f_{\sigma}(\tau) + N S \lambda + \frac{N}{J_{K}} b_{0}^{\dagger}(\tau) b_{0}(\tau) \Big\} \nn &&
+ d\Lambda\sum_{w = 1}^{f} \int_{0}^{\beta} d \tau \Big\{ \frac{\pi N \Lambda_b}{J_{K} (d \Lambda)^2} \Big( b_{w}^{\dagger}(\tau) - b_{w-1}^{\dagger}(\tau) \Big) \Big( b_{w}(\tau) - b_{w-1}(\tau) \Big) + \frac{2 N S N_{F}}{v_{F} \Lambda_{c}^{2}} b_{w-1}(\tau) \partial_{\tau} b_{w-1}^{\dagger}(\tau) \Big\} \bigg] . \eqa
\end{widetext}
As clarified in subsection \ref{dz_Equal_dLambda}, $d\Lambda \sum_{w=1}^f$ corresponds to $\int_0^{z_f}\!dz$ .

Taking the $f = 0$ limit, it is straightforward to see that this effective field theory is reduced to the mean-field theory
\bqa && Z = Z_{c} \int D f_{\sigma}(\tau) \exp \bigg[ - \int_{0}^{\beta} d \tau \Big\{ \int_{0}^{\beta} d \tau' f_{\sigma}^{\dagger}(\tau) b_{0}(\tau) \nn && \times G_{c}(\tau-\tau') b_{0}^{\dagger}(\tau') f_{\sigma}(\tau') + f_{\sigma}^{\dagger}(\tau) (\partial_{\tau} - \lambda) f_{\sigma}(\tau) \nn && + N S \lambda + \frac{N}{J_{K}} b_{0}^{\dagger}(\tau) b_{0}(\tau) \Big\} \bigg] , \nonumber \eqa
Performing the Gaussian integration for the fermion variable, we obtain an effective partition function in the large-$N$ limit
\bqa && Z = Z_{c} \exp\Big[ N \mbox{tr}_{\tau\tau'} \ln \Big( (\partial_{\tau} - \lambda) \delta(\tau-\tau') \nn && + b_{0}(\tau) G_{c}(\tau-\tau') b_{0}^{\dagger}(\tau') \Big) - \int_{0}^{\beta} d \tau \Big\{ N S \lambda \nn && + \frac{N}{J_{K}} b_{0}^{\dagger}(\tau) b_{0}(\tau) \Big\} \Big] \equiv \exp\Big( - \beta F[b_{0}(\tau)] \Big) . \eqa
Here, $F[b_{0}(\tau)]$ is the mean-field free energy for the Kondo effect. It is easy to find the mean-field equation for the hybridization order parameter \cite{Kondo_Textbook}.

\subsubsection{Quantum corrections up to the $1/N$ order}

Now, we take $f = 1$. Introducing
\bqa && b_{1}(\tau) = b_{0}(\tau) + \delta b_{1}(\tau) \eqa
into the effective Landau-Ginzburg field theory, we obtain
\begin{widetext}
\bqa && Z = Z_{c} Z_{h} \int D f_{\sigma}(\tau) D \delta b_{1}(\tau) \exp\Big[ - \int_{0}^{\beta} d \tau \Big\{ \int_{0}^{\beta} d \tau' f_{\sigma}^{\dagger}(\tau) \Big( b_{0}(\tau) + \delta b_{1}(\tau) \Big) G_{c}(\tau-\tau') \Big( b_{0}^{\dagger}(\tau') + \delta b_{1}^{\dagger}(\tau') \Big) f_{\sigma}(\tau') \nn && + f_{\sigma}^{\dagger}(\tau) (\partial_{\tau} - \lambda) f_{\sigma}(\tau) + N S \lambda + \frac{N}{J_{K}} b_{0}^{\dagger}(\tau) b_{0}(\tau) + \frac{\pi N \Lambda_b}{J_{K} d \Lambda} \delta b_{1}^{\dagger}(\tau) \delta b_{1}(\tau) + d \Lambda \frac{2 N S N_{F}}{v_{F} \Lambda_{c}^{2}} b_{0}(\tau) \partial_{\tau} b_{0}^{\dagger}(\tau) \Big\} \Big] . \eqa
\end{widetext}
This effective partition function gives rise to coupled equations for both hybridization order parameters of $b_{0}(\tau)$ and $\delta b_{1}(\tau)$. Solving the equation for $\delta b_{1}(\tau)$, one represents it as a function of $b_{0}(\tau)$. Inserting this expression into the equation for $b_{0}(\tau)$, we obtain an equation for the hybridization order parameter $b_{0}(\tau)$ with the introduction of quantum corrections in the $1/N$ expansion.

Equivalently, we perform the Gaussian integration for the $\delta b_{1}(\tau)$ variable, and obtain
\begin{widetext}
\bqa && Z = Z_{c} Z_{h} \int D f_{\sigma}(\tau) \exp\Big[ - \int_{0}^{\beta} d \tau \Big\{ \int_{0}^{\beta} d \tau' f_{\sigma}^{\dagger}(\tau) b_{0}(\tau) G_{c}(\tau-\tau') b_{0}^{\dagger}(\tau') f_{\sigma}(\tau') + f_{\sigma}^{\dagger}(\tau) (\partial_{\tau} - \lambda) f_{\sigma}(\tau) + N S \lambda \nn && + \frac{N}{J_{K}} b_{0}^{\dagger}(\tau) b_{0}(\tau) + d \Lambda \frac{2 N S N_{F}}{v_{F} \Lambda_{c}^{2}} b_{0}(\tau) \partial_{\tau} b_{0}^{\dagger}(\tau) \Big\} - \mbox{tr}_{\tau\tau'} \ln \Big( \frac{\pi N \Lambda_b}{J_{K} d \Lambda} \delta(\tau-\tau') + f_{\sigma}^{\dagger}(\tau) G_{c}(\tau-\tau') f_{\sigma}(\tau') \Big) \nn && + \int_{0}^{\beta} d \tau \int_{0}^{\beta} d \tau' \int_{0}^{\beta} d \tau'' \int_{0}^{\beta} d \tau''' \Big\{ f_{\sigma}^{\dagger}(\tau) G_{c}(\tau-\tau'') b_{0}^{\dagger}(\tau'') f_{\sigma}(\tau'') \Big( \frac{\pi N\Lambda_b}{J_{K} d \Lambda} \delta(\tau-\tau') \nn && + f_{\sigma''}^{\dagger}(\tau) G_{c}(\tau-\tau') f_{\sigma''}(\tau') \Big)^{-1} f_{\sigma'}^{\dagger}(\tau''') b_{0}(\tau''') G_{c}(\tau'''-\tau') f_{\sigma'}(\tau') \Big\} \Big] . \eqa
\end{widetext}

Finally, the Gaussian integration of the fermion variable results in the effective action
\bqa && \mathcal{S}_{eff} = \mathcal{S}_{N \rightarrow \infty} + \mathcal{S}_{1/N} + \Delta \mathcal{S}_{1/N} . \eqa
Here,
\bqa && \mathcal{S}_{N \rightarrow \infty} = - N \mbox{tr}_{\tau\tau'} \ln \Big( (\partial_{\tau} - \lambda) \delta(\tau - \tau') \nn && + b_{0}(\tau) G_{c}(\tau-\tau') b_{0}^{\dagger}(\tau') \Big) + \int_{0}^{\beta} d \tau \Big\{ N S \lambda \nn && + \frac{N}{J_{K}} b_{0}^{\dagger}(\tau) b_{0}(\tau) + d \Lambda \frac{2 N S N_{F}}{v_{F} \Lambda_{c}^{2}} b_{0}(\tau) \partial_{\tau} b_{0}^{\dagger}(\tau) \Big\} \eqa
is the effective action in the large$-N$ limit, which results from the integration of the spinon variable and gives rise to the mean-field equation.
\bqa \mathcal{S}_{1/N} &=& \mbox{tr}_{\tau\tau'} \ln \Big( \frac{\pi N \Lambda_b}{J_{K} d \Lambda} \delta(\tau-\tau') \nn &+& N G_{c}(\tau-\tau') G_{f}(\tau'-\tau) \Big) \eqa
is the $1/N$ correction for the effective free energy, which comes from the integration of the fluctuating hybridization order parameter $\delta b_{1}(\tau)$ and leads the hybridization order parameter $b_{0}(\tau)$ to vanish beyond the mean-field approximation \cite{Kondo_1_N_Expansion}. The last part of
\bqa && \Delta \mathcal{S}_{1/N} = \int_{0}^{\beta} d \tau \int_{0}^{\beta} d \tau' \int_{0}^{\beta} d \tau'' \int_{0}^{\beta} d \tau''' \nn && \Big\{ G_{c}(\tau-\tau'') b_{0}^{\dagger}(\tau'') G_{f}(\tau''-\tau''') \Big( \frac{\pi\Lambda_b}{J_{K} d \Lambda} \delta(\tau-\tau') \nn && + G_{c}(\tau-\tau') G_{f}(\tau'-\tau) \Big)^{-1} b_{0}(\tau''') G_{c}(\tau'''-\tau') \nn && \times G_{f}(\tau'-\tau) \Big\} \eqa
describes a shift of the order parameter from the original mean-field value in the presence of the $1/N$ correction. It should appear with self-consistency. However, this shift term is not taken into account usually for the analysis without self-consistency, which is also in the order of $1/N$.

\subsubsection{Quantum corrections up to the $1/N^{2}$ order}

We consider $f = 2$. Introducing
\bqa b_{1}(\tau) &=& b_{0}(\tau) + \delta b_{1}(\tau) , \nn b_{2}(\tau) &=& b_{1}(\tau) + \delta b_{2}(\tau) \nn &=& b_{0}(\tau) + \delta b_{1}(\tau) + \delta b_{2}(\tau) \eqa
into the effective Landau-Ginzburg field theory, we obtain
\begin{widetext}
\bqa && Z = Z_{c} Z_{h}^{2} \int D f_{\sigma}(\tau) D \delta b_{1}(\tau) D \delta b_{2}(\tau) \exp\Big[ - \int_{0}^{\beta} d \tau \Big\{ \int_{0}^{\beta} d \tau' f_{\sigma}^{\dagger}(\tau) \Big( b_{0}(\tau) + \delta b_{1}(\tau) + \delta b_{2}(\tau) \Big) G_{c}(\tau-\tau') \Big( b_{0}^{\dagger}(\tau') \nn && + \delta b_{1}^{\dagger}(\tau') + \delta b_{2}^{\dagger}(\tau') \Big) f_{\sigma}(\tau') + f_{\sigma}^{\dagger}(\tau) (\partial_{\tau} - \lambda) f_{\sigma}(\tau) + N S \lambda + \frac{N}{J_{K}} b_{0}^{\dagger}(\tau) b_{0}(\tau) + \frac{\pi N \Lambda_b}{J_{K} d \Lambda} \delta b_{1}^{\dagger}(\tau) \delta b_{1}(\tau) \nn && + \frac{\pi  N \Lambda_b}{J_{K} d \Lambda} \delta b_{2}^{\dagger}(\tau) \delta b_{2}(\tau) + d \Lambda \frac{2 N S N_{F}}{v_{F} \Lambda_{c}^{2}} b_{0}(\tau) \partial_{\tau} b_{0}^{\dagger}(\tau) + d \Lambda \frac{2 N S N_{F}}{v_{F} \Lambda_{c}^{2}} \Big( b_{0}(\tau) + \delta b_{1}(\tau) \Big) \Big( \partial_{\tau} b_{0}^{\dagger}(\tau) + \partial_{\tau} \delta b_{1}^{\dagger}(\tau) \Big) \Big\} \Big] . \eqa
\end{widetext}
Now, we have three coupled equations for $b_{0}(\tau)$, $\delta b_{1}(\tau)$, and $\delta b_{2}(\tau)$.

Performing Gaussian integrals for $\delta b_{1}(\tau)$ and $\delta b_{2}(\tau)$, we obtain an effective action
\bqa && \mathcal{S}_{eff} = \mathcal{S}_{N \rightarrow \infty} + \mathcal{S}_{1/N} + \mathcal{S}_{1/N^{2}} + \Delta \mathcal{S}_{1/N} + \Delta \mathcal{S}_{1/N^{2}} . \nn \eqa
Here,
\bqa && \mathcal{S}_{1/N^{2}} = \mbox{tr}_{\tau\tau'} \ln \Big\{ \frac{\pi N \Lambda_b}{J_{K} d \Lambda} \delta(\tau-\tau') \nn && + N G_{c}(\tau-\tau') G_{f}(\tau'-\tau) + d \Lambda \frac{2 N S N_{F}}{v_{F} \Lambda_{c}^{2}} \partial_{\tau} \delta(\tau-\tau') \nn && - \int_{0}^{\beta} d \tau'' \int_{0}^{\beta} d \tau''' G_{c}(\tau-\tau'') G_{f}(\tau''-\tau''') \nn && \times \Big(\frac{\pi\Lambda_b}{J_{K} d \Lambda} \delta(\tau'''-\tau'') +  G_{c}(\tau'''-\tau'') G_{f}(\tau''-\tau''') \Big)^{-1} \nn && \times G_{c}(\tau'''-\tau') G_{f}(\tau'-\tau) \Big\} \eqa
is the $1/N^{2}$ correction in the effective action, in particular, the last term in the logarithm. $\Delta \mathcal{S}_{1/N^{2}}$ is a shift term in the $1/N^{2}$ order, which will not be shown here due to its complex expression.

\subsubsection{A non-perturbative large $N$ effective field theory}

It is remarkable to obtain a non-perturbative large $N$ effective field theory with an emergent extra dimension, summing all loop quantum corrections organized in the $1/N$ expansion. Frankly speaking, we do not understand why this non-perturbative large $N$ effective field theory with an emergent extra dimension should arise from our recursive Wilsonian renormalization group transformations. Furthermore, we do not have a clear physical picture on how this recursive Wilsonian renormalization group transformation method takes into account such non-perturbative physical effects. It is necessary to understand the mathematical structure more deeply.

\subsection{Comparison between the Wilson's numerical renormalization group approach and our recursive Wilsonian renormalization group method}

One cautious person may point out several correspondences between the Wilson's numerical renormalization group approach and our recursive Wilsonian renormalization group method. Actually, we believe that our recursive Wilsonian renormalization group method can be regarded to be one mathematical reformulation of the Wilson's numerical renormalization group analysis in the path integral representation.

The Wilson's numerical renormalization group analysis consists of (a) Division of the energy support of the bath spectral function into a set of logarithmic intervals, (b) Reduction of the continuous spectrum to a discrete
set of states (logarithmic discretization), (c) Mapping of the discretized model onto a semi-infinite chain, where an impurity lies at one end and effective bath fermions appear in the other side, (d) Iterative diagonalization of this chain, increasing the size of the chain and attaching bath fermions into the original chain, and (e) Further analysis of the many-particle energies, matrix elements, etc., calculated during the iterative diagonalization, which yields information on fixed points and static and dynamic properties of the quantum impurity model \cite{Bulla_RMP}. On the other hand, recursive Wilsonian renormalization group transformations are performed in the following way. (1) We introduce an order parameter field to diagonalize the corresponding effective Hamiltonian in the saddle point approximation. (2) We separate slow and fast degrees of freedom for all variables, including order parameter fields. (3) Integrating over fast degrees of freedom to renormalize the dynamics of slow degrees of freedom and rescaling both the UV cutoff to return to its original value and all slow functional variables to make the resulting effective action invariant, we finish the first iteration procedure of the Wilsonian renormalization group analysis, where effective interactions still exist for slow degrees of freedom. (4) To deal with these effective interactions, we perform the Hubbard-Stratonovich transformation once again and obtain an effective theory with essentially the same order parameter field but in the second iteration step. (5) We repeat exactly the same procedures of (2) and (3) and find an effective theory in terms of slower degrees of freedom after the second iteration procedure. This procedure gives rise to recursive renormalization group transformations, where a renormalized (hybridization) order parameter in the first depth is utilized to renormalize the (hybridization) order parameter in the second depth. This recursive structure of renormalization group transformations is responsible for the introduction of all-loop quantum corrections, organized in the $1/N$ expansion.

Now, let's make correspondences between the Wilson's numerical renormalization group approach and the recursive Wilsonian renormalization group method. The logarithmic discretization procedure (a) and (b) may be identified with the separation of slow and fast degrees of freedom with a logarithmic cutoff (2), regarded to be a typical procedure of the Wilsonian renormalization group transformation. (c) Mapping of the discretized model onto a semi-infinite chain corresponds to (1) $\&$ (4) the introduction of an order parameter field through the Hubbard-Stratonovich transformation, regarded to be a way for matrix diagonalization. Although both (c) and (1) $\&$ (4) are important in the technical aspect, they are not essential in the view of principle. A crucial point is on the correspondence between (d) Iterative diagonalization of this semi-infinite chain model and (3) $\&$ (5) recursive Wilsonian renormalization group transformations, which is the core of the renormalization group transformation.

In the Wilson's numerical renormalization group approach, one diagonalizes the effective semi-infinite chain Hamiltonian with the size $N$, where an impurity lies at one end and effective bath fermions appear in the other side. Then, one attaches effective bath fermions into the chain Hamiltonian of the size $N$ at the $N+1$ site, where an effective hopping integral is reduced in a logarithmic way and determined by the mapping procedure of the discretized model onto a semi-infinite chain. Now, one can diagonalize the $N+1$ effective chain Hamiltonian and obtain the flow of the eigenvalue as a function of the size $N$. During this diagonalization procedure, one may perform truncation to reduce the size of the total Hilbert space for the exact diagonalization. We argue that this numerical renormalization group structure is in parallel with the present recursive renormalization group transformation. In particular, the renormalized (hybridization) order parameter in the $(f-1)$th iteration is utilized to renormalize the (hybridization) order parameter in the $f$th iteration, where the exact diagonalization procedure of the numerical renormalization group analysis corresponds to the integration of high-energy modes of all field variables in the $d \Lambda \rightarrow 0$ limit. In this respect the size of the effective chain Hamiltonian may be identified with the emergent extra dimension of the recursive renormalization group method.

Unfortunately, the above discussion is qualitative or speculative. To make the correspondence more quantitatively, we need to construct recursion equations for two successive eigenvalues and eigenstates between the $N$th and $(N+1)$th effective chain Hamiltonians in the numerical renormalization group structure. Then, we may perform the continuation of the size variable and obtain non-perturbative renormalization group flows in the numerical renormalization group method. One may find a structure of the holographic duality conjecture in these recursion equations along the emergent extra dimension, compared with the present theoretical framework. This would be an interesting future direction of research.

\section{The Kondo effect in the non-perturbative field theoretical framework} \label{Solution_Kodno_Effect}

The mean-field theory for the Kondo effect describes the local-Fermi liquid fixed point well, where the hybridization order parameter becomes finite below the Kondo temperature \cite{Kondo_Textbook}. On the other hand, it gives rise to a second order phase transition, which should be regarded as an artifact of the mean-field theory. Introducing quantum corrections at the leading order of the $1/N$ expansion into the mean-field theory, the hybridization order parameter turns out to vanish for all temperatures, where the nature of the local Fermi-liquid fixed point is not perfectly clarified \cite{Kondo_1_N_Expansion}.

We claim that this non-perturbative physics would be encoded into our effective Landau-Ginzburg description for the Kondo effect beyond the $1/N$ framework, where the evolution of the hybridization order parameter is given by the diffusion equation in the emergent spacetime with an extra dimension. Given the UV boundary condition Eq. (\ref{UV_Boundary_Condition}), renormalization of the hybridization order parameter occurs through the diffusion equation Eq. (\ref{Diffusion_Equation}). Finally, the diffusive evolution matches or is constrained by the IR boundary condition Eq. (\ref{IR_Boundary_Condition}). In particular, the diffusive dynamics along the direction of the extra dimension would erase the memory of the emergent low-energy degrees of freedom, giving rise to short-range correlations in time. It turns out that this scale determines the Kondo temperature.

\subsection{Solution for the non-perturbative renormalization group equation: Hybridization order parameter}

We recall Eq. (\ref{Diffusion_Equation})
\bqa && \partial_{\tau} b^{\dagger}(\tau,z) = D \partial_{z}^{2} b^{\dagger}(\tau,z) , \nonumber \eqa
where
\bqa && D \equiv \frac{1}{NS\tilde g_c\tilde g_h^b} = \frac{\pi v_{F}  \Lambda_b \Lambda_c^2}{2 S N_{F} J_{K}} \eqa
may be identified with an effective diffusion constant.

We consider
\bqa && b^{\dagger}(\tau,z) = e^{- C \tau} \Big\{ A \cos \Big( \sqrt{\frac{C}{D}} z \Big) + B \sin \Big( \sqrt{\frac{C}{D}} z \Big) \Big\} \nn \eqa
as a trial solution of this effective diffusion equation. $A$, $B$, and $C$ will be determined by both UV and IR boundary conditions.

The UV boundary condition Eq. (\ref{UV_Boundary_Condition}) fixes the relation between $A$ and $B$. As a result, we obtain
\bqa && b^{\dagger}(\tau,z) = B e^{- C \tau} \Big\{ \alpha\sqrt{\frac{C}{D}} \cos \Big( \sqrt{\frac{C}{D}} z \Big) + \sin \Big( \sqrt{\frac{C}{D}} z \Big) \Big\} , \nn \label{Nonperturbative_Solution} \eqa
where $\alpha = J_K / (N \tilde g_h^b) = \pi \Lambda_b$.

Applying the IR boundary condition Eq. (\ref{IR_Boundary_Condition}) into this solution, we obtain
\bqa && \Big\{ - \alpha \frac{C}{D} \sin \Big( \sqrt{\frac{C}{D}} z_{f} \Big) + \sqrt{\frac{C}{D}} \cos \Big( \sqrt{\frac{C}{D}} z_{f} \Big) \Big\} \nn && + N\tilde g_h^b \Big\{ \alpha \sqrt{\frac{C}{D}} \cos \Big( \sqrt{\frac{C}{D}} z_{f} \Big) + \sin \Big( \sqrt{\frac{C}{D}} z_{f} \Big) \Big\} \nn && \times \frac{1}{\beta} \sum_{i \omega} G_{c}(i \omega ) G_{f}(i \omega - C) = 0 . \label{IR_Boundary_Condition_Solution} \eqa

Performing the Fourier transformation in Eq. (\ref{Spinon_Green_Function}), we find
\bqa && G_{f}(i \omega) = \Big\{ i \omega + \lambda \nn && - B^{2} \Big\{ \alpha \sqrt{\frac{C}{D}} \cos \Big( \sqrt{\frac{C}{D}} z_{f} \Big) + \sin \Big( \sqrt{\frac{C}{D}} z_{f} \Big) \Big\}^{2} G_{c}(i \omega + C) \Big\}^{-1} . \nn \eqa
Inserting the spinon Green's function into Eq. (\ref{IR_Boundary_Condition_Solution}), we obtain
\begin{widetext}
\bqa && \Big\{ - \alpha\frac{C}{D} \sin \Big( \sqrt{\frac{C}{D}} z_{f} \Big) + \sqrt{\frac{C}{D}} \cos \Big( \sqrt{\frac{C}{D}} z_{f} \Big) \Big\} + N \tilde g_h^b \Big\{ \alpha \sqrt{\frac{C}{D}} \cos \Big( \sqrt{\frac{C}{D}} z_{f} \Big) + \sin \Big( \sqrt{\frac{C}{D}} z_{f} \Big) \Big\} \nn &&\hspace{100pt} \times \frac{1}{\beta} \sum_{i \omega} \frac{ - i \pi N_{F} \mbox{sign}(\omega) }{i \omega + \lambda - C + i \pi N_{F} B^{2} \Big\{ \alpha \sqrt{\frac{C}{D}} \cos \Big( \sqrt{\frac{C}{D}} z_{f} \Big) + \sin \Big( \sqrt{\frac{C}{D}} z_{f} \Big) \Big\}^{2} \mbox{sign}(\omega)} = 0 , \label{Self_Consistent_Equation_C} \eqa
\end{widetext}
which determines $C$. Below, we show that the amplitude $B$ of the hybridization order parameter is scaled out. The constraint equation is given by
\begin{widetext}
\bqa && \frac{1}{\beta} \sum_{i \omega} \frac{1}{i \omega + \lambda - C + i \pi N_{F} B^{2} \Big\{ \alpha \sqrt{\frac{C}{D}} \cos \Big( \sqrt{\frac{C}{D}} z_{f} \Big) + \sin \Big( \sqrt{\frac{C}{D}} z_{f} \Big) \Big\}^{2} \mbox{sign}(\omega)} = S , \eqa
\end{widetext}
which determines $\lambda$. We recall $S = 1/2$.

Introducing $N_{F} \rightarrow \frac{N_{F}}{B^{2}}$ and $J_{K} \rightarrow J_{K} B^{2}$ into these coupled equations, one can show that $B$ disappears as follows
\begin{widetext}
\bqa && \Big\{ - \alpha \frac{C}{D} \sin \Big( \sqrt{\frac{C}{D}} z_{f} \Big) + \sqrt{\frac{C}{D}} \cos \Big( \sqrt{\frac{C}{D}} z_{f} \Big) \Big\} + N \tilde g_h^b \Big\{ \alpha \sqrt{\frac{C}{D}} \cos \Big( \sqrt{\frac{C}{D}} z_{f} \Big) + \sin \Big( \sqrt{\frac{C}{D}} z_{f} \Big) \Big\} \nn && \hspace{100pt} \times \frac{1}{\beta} \sum_{i \omega} \frac{ - i \pi N_{F} \mbox{sign}(\omega) }{i \omega + \lambda - C + i \pi N_{F} \Big\{ \alpha \sqrt{\frac{C}{D}} \cos \Big( \sqrt{\frac{C}{D}} z_{f} \Big) + \sin \Big( \sqrt{\frac{C}{D}} z_{f} \Big) \Big\}^{2} \mbox{sign}(\omega)} = 0 , \nn && \frac{1}{\beta} \sum_{i \omega} \frac{1}{i \omega + \lambda - C + i \pi N_{F} \Big\{ \alpha \sqrt{\frac{C}{D}} \cos \Big( \sqrt{\frac{C}{D}} z_{f} \Big) + \sin \Big( \sqrt{\frac{C}{D}} z_{f} \Big) \Big\}^{2} \mbox{sign}(\omega)} = S . \label{Self_Consistent_Equations} \eqa
\end{widetext}
These coupled equations describe the Kondo effect in a non-perturbative way.

The second equation of Eq. (\ref{Self_Consistent_Equations}) leads to $\lambda = C$ for all temperatures. This is due to the condition of half filling with particle-hole symmetry. Inserting this relation into the first equation of Eq. (\ref{Self_Consistent_Equations}), we obtain
\begin{eqnarray}
\frac{1}{ N \tilde g_h^b} \frac{- \alpha \frac{C}{D} \sin \Big( \sqrt{\frac{C}{D}} z_{f} \Big) + \sqrt{\frac{C}{D}} \cos \Big( \sqrt{\frac{C}{D}} z_{f} \Big)}{\alpha \sqrt{\frac{C}{D}} \cos \Big( \sqrt{\frac{C}{D}} z_{f} \Big) + \sin \Big( \sqrt{\frac{C}{D}} z_{f} \Big)} = I
\label{eq:SCE_C}
\end{eqnarray}
with
\begin{eqnarray}
&&I = \frac{1}{\beta} \sum_{i \omega} i \pi N_{F}{\rm sign}(\omega) \bigg[ i \omega + i \pi N_{F} {\rm sign}(\omega) \nn
&& \times \Big\{ \alpha \sqrt{\frac{C}{D}} \cos \Big( \sqrt{\frac{C}{D}} z_{f} \Big) + \sin \Big( \sqrt{\frac{C}{D}} z_{f} \Big) \Big\}^{2} \bigg]^{-1} .
\end{eqnarray}

In order to solve this order-parameter equation at finite temperatures, we set the cutoff $z_f = f d\Lambda$ as $z_{f} = \Lambda - T$ with $\Lambda \gg T$. The renormalization group procedure should be terminated at a given temperature. This finite-temperature termination is introduced into the cutoff phenomenologically. However, it turns out that the reduction effect is not serious as long as $\Lambda \gg T$.

\begin{figure}[t!]
\includegraphics[width=8cm]{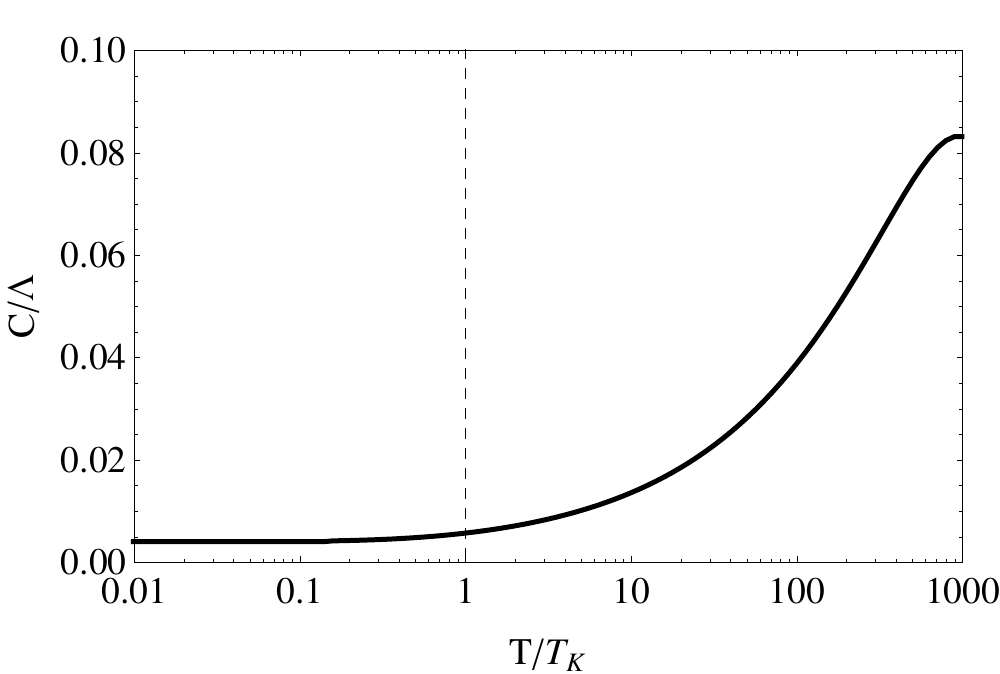}
\caption{ Linear-log plot of the order parameter $C/\Lambda$ in terms of $T/T_{K}$ for $J_K\Lambda = 1.05$ and $N_F/\Lambda = 0.1$. Here, $T_{K}$ is an effective Kondo temperature, introduced from our field theoretical framework. $C$ is nearly constant at low temperatures less than the Kondo temperature and gradually increases above the Kondo temperature.
}
\label{Fig:C}
\end{figure}

Solving this equation, we obtain $C$ as a function of temperature as shown in Fig. \ref{Fig:C}, where the temperature axis is scaled by the Kondo temperature. Considering the self-energy correction in the spinon's Green's function, it is natural to take the Kondo temperature as
%
%
\begin{eqnarray}
T_{K} = \pi N_{F} \Big\{ \alpha \sqrt{\frac{C_0}{D}} \cos \Big( \sqrt{\frac{C_0}{D}} \Big) + \sin \Big( \sqrt{\frac{C_0}{D}} \Big) \Big\}^{2},
\end{eqnarray}
where $C_0$ is the solution of Eq. (\ref{eq:SCE_C}) at $T=0$. Figure \ref{Fig:C} shows that $C$ is nearly constant at low temperatures less than the Kondo temperature and gradually increases above the Kondo temperature.

\subsection{Impurity thermodynamics}

Performing the Gaussian integration for the spinon variable in the effective Landau-Ginzburg field theory Eq. (\ref{Effective_Landau_Ginzburg_Theory}) and introducing the non-perturbative solution Eq. (\ref{Nonperturbative_Solution}) of the hybridization order parameter into the effective Landau-Ginzburg field theory Eq. (\ref{Effective_Landau_Ginzburg_Theory}), we find the free energy functional for the Kondo effect as follows
\begin{widetext}
\bqa && F = - z_{f} \frac{N}{\beta} \ln Z_{h} - \frac{N}{\beta} \ln Z_{c} - \frac{N}{\beta} \sum_{i \omega} \ln \Big\{ - i \omega - \lambda + C - i \pi N_{F} \Big\{ \alpha \sqrt{\frac{C}{D}} \cos \Big( \sqrt{\frac{C}{D}} z_{f} \Big) + \sin \Big( \sqrt{\frac{C}{D}} z_{f} \Big) \Big\}^{2} \mbox{sign}(\omega) \Big\} \nn && + N S \lambda + {NS\over \beta} {N_F\over v_F\Lambda_c^2} \Big( 1 - e^{-\beta C} \Big) \bigg\{ {\alpha^2\over2} \sqrt{C\over D} \sin \Big( 2\sqrt{C\over D}z_f \Big) - {1\over2} \sqrt{D\over C} \sin \Big( 2\sqrt{C\over D} z_f \Big) + {\alpha\over2} - {\alpha\over2} \cos \Big( 2\sqrt{C\over D} z_f \Big) \bigg\} . \label{Free_Energy_Full_Expression} \eqa
\end{widetext}
Here, the hybridization order parameter field is determined by its self-consistent equations [Eq. (\ref{Self_Consistent_Equations})].

In order to figure out thermodynamics of the Kondo effect, we focus on the most singular sector, given by
\bqa && F_{imp} = - \frac{N}{\beta} \sum_{i \omega} \ln \bigg[ - i \omega - \lambda + C - i\pi N_F {\rm sign}(\omega) \nn && \times \Big\{ \alpha\sqrt{\frac{C}{D}} \cos \Big( \sqrt{\frac{C}{D}} z_{f} \Big) + \sin \Big( \sqrt{\frac{C}{D}} z_{f} \Big) \Big\}^{2} \bigg] \nn && + N S \lambda , \nonumber \eqa
which is nothing but Eq. (\ref{Impurity_Thermodynamics_Singular_Part}). This expression implies that the effective Kondo temperature is
\bqa T_{K} = \pi N_{F} \Big\{ \alpha\sqrt{\frac{C_0}{D}} \cos \Big( \sqrt{\frac{C_0}{D}} \Big) + \sin \Big( \sqrt{\frac{C_0}{D}} \Big) \Big\}^{2}, \nonumber \eqa
as discussed earlier. Based on this effective free energy functional, we describe non-perturbative thermodynamics in this exactly screened Kondo problem, as discussed before.

\section{Implication for the holographic duality conjecture: How to find an emergent metric structure}

\subsection{Hamilton-Jacobi formulation: Induced boundary metric}

Finally, we discuss how to extract out the emergent metric structure. We recall the partition function
\begin{widetext}
\bqa && W_{IR}(z_{f}) = \int D f_{\sigma}(\tau) D b(\tau,z) \exp\bigg[ - \int_{0}^{z_{f}} d z \int_{0}^{\infty} d \tau \bigg\{ \frac{\pi N\Lambda_b}{J_{K}} \Big( \partial_{z} b^{\dagger}(\tau,z) \Big) \Big( \partial_{z} b(\tau,z) \Big) + \frac{2 N S N_{F}}{v_{F} \Lambda_{c}^{2}} b(\tau,z) \partial_{\tau} b^{\dagger}(\tau,z) \bigg\} \nn && - \int_{0}^{\infty} d \tau \bigg\{ \int_{0}^{\infty} d \tau' f_{\sigma}^{\dagger}(\tau) b(\tau,z_{f}) G_{c}(\tau-\tau') b^{\dagger}(\tau',z_{f}) f_{\sigma}(\tau') + f_{\sigma}^{\dagger}(\tau) (\partial_{\tau} - \lambda ) f_{\sigma}(\tau) + N S \lambda + \frac{N}{J_{K}} b^{\dagger}(\tau,0) b(\tau,0) \bigg\} \bigg] . \label{Effective_Free_Energy_Extra_Dimension} \eqa
\end{widetext}

We point out that this partition function should not depend on the scale of $z_{f}$, i.e.,
\bqa && \frac{d }{d z_{f}} \ln W_{IR}(z_{f}) = 0 . \label{Callan_Symanzik_Equation} \eqa
As a result, we obtain the following equation
\begin{widetext}
\bqa && 0 = \int_{0}^{\infty} d \tau \bigg\{ \frac{\pi N \Lambda_b}{J_{K}} \Big( \partial_{f} b^{\dagger}(\tau,z_{f}) \Big) \Big( \partial_{f} b(\tau,z_{f}) \Big) + \frac{2 N S N_{F}}{v_{F} \Lambda_{c}^{2}} b(\tau,z_{f}) \partial_{\tau} b^{\dagger}(\tau,z_{f}) \nn && + N \int_{0}^{\infty} d \tau' \partial_{f} b(\tau,z_{f}) G_{c}(\tau-\tau') b^{\dagger}(\tau',z_{f}) G_{f}(\tau'-\tau) + N \int_{0}^{\infty} d \tau' b(\tau,z_{f}) G_{c}(\tau-\tau') \partial_{f} b^{\dagger}(\tau',z_{f}) G_{f}(\tau'-\tau) \bigg\} . \eqa
\end{widetext}
Here, $\partial_{f} b(\tau,z_{f})$ is a short-handed notation for $\partial_{z} b(\tau,z) \big|_{z = z_{f}}$. This is nothing but the Hamilton-Jacobi equation.

The Hamilton-Jacobi formulation consists of three types of equations \cite{deBoer:1999tgo,Verlinde:1999xm}. The first are equations that counter terms should satisfy, the second is the Hamilton-Jacobi equation, and the third are equations to define the energy-momentum tensor. We emphasize that the Hamilton-Jacobi equation is just a reformulation of our equations of motion, that is, the UV boundary condition, the bulk equation of motion, and the IR boundary condition, which will be clairified below.  The only thing that does not seem to take into account is the issue on counter terms. Since we are dealing with an effective field theory defined in the cutoffs of $v_{F} \Lambda_{c} \gg \Lambda_{f}, ~ \Lambda_{b}$, there do not exist any divergences, which reflects the procedure of the Wilsonian renormalization group structure.

The Callan-Symanzik equation (\ref{Callan_Symanzik_Equation}) can be rewritten as the Hamilton-Jacobi equation \cite{Bianchi:2001kw,deBoer:1999tgo,Verlinde:1999xm,AdS_CFT_Review}
\begin{equation}
\mathcal{H} + \partial_{f} \mathcal{S}_{IR} = 0 . \nonumber
\end{equation}
The bulk ``Hamiltonian" is identified with
\bqa
&& \mathcal{H} = \int_{0}^{\infty} d \tau \Big\{ \frac{\pi N\Lambda_b}{J_{K}} \Big( \partial_{f} b^{\dagger}(\tau,z_{f}) \Big) \Big( \partial_{f} b(\tau,z_{f}) \Big) \nn && + \frac{2 N S N_{F}}{v_{F} \Lambda_{c}^{2}} b(\tau,z_{f}) \partial_{\tau} b^{\dagger}(\tau,z_{f}) \Big\} , \nonumber
\eqa
and the IR effective action is given by
\bqa
&& \mathcal{S}_{IR} = \int_{0}^{\infty} d \tau \Big\{ f_{\sigma}^{\dagger}(\tau) \partial_{\tau} f_{\sigma}(\tau) \nn && + \int_{0}^{\infty} d \tau' f_{\sigma}^{\dagger}(\tau) b(\tau,z_{f}) G_{c}(\tau-\tau') b^{\dagger}(\tau',z_{f}) f_{\sigma}(\tau') \Big\} , \nonumber
\eqa
which defines the energy-momentum tensor. This IR effective action is also discussed below.

Recalling the IR boundary condition given by Eq. (\ref{IR_Boundary_Condition}), the Hamilton-Jacobi equation can be more simplified as follows
\bqa && 0 = \int_{0}^{\infty} d \tau \bigg\{ \frac{2 N S N_{F}}{v_{F} \Lambda_{c}^{2}} b(\tau,z_{f}) \partial_{\tau} b^{\dagger}(\tau,z_{f}) \nn && + N b(\tau,z_{f}) \int_{0}^{\infty} d \tau' G_{c}(\tau-\tau') \partial_{f} b^{\dagger}(\tau',z_{f}) G_{f}(\tau'-\tau) \bigg\} . \nn \label{RG_Equation_from_GR} \eqa
We would like to emphasize that this renormalization group equation can be regarded as one reformulation of the IR boundary condition [Eq. (\ref{IR_Boundary_Condition})] with the bulk equation of motion [Eq. (\ref{Diffusion_Equation})].

Applying $\partial_{f}$ into the IR boundary condition [Eq. (\ref{IR_Boundary_Condition})], we obtain
\bqa && \frac{\pi\Lambda_b}{J_{K}} \partial_{f}^{2} b^{\dagger}(\tau,z_{f}) \nn && + \int_{0}^{\infty} d \tau' G_{c}(\tau-\tau') \partial_{f} b^{\dagger}(\tau',z_{f}) G_{f}(\tau' - \tau) = 0 . \nonumber \eqa
Replacing $\partial_{f}^{2} b^{\dagger}(\tau,z_{f})$ with a time$-$derivative term via the bulk equation of motion [Eq. (\ref{Diffusion_Equation})], we find
\bqa && \frac{2 S N_{F}}{v_{F} \Lambda_{c}^{2}} b(\tau,z_{f}) \partial_{\tau} b^{\dagger}(\tau,z_{f}) \nn && + b(\tau,z_{f}) \int_{0}^{\infty} d \tau' G_{c}(\tau-\tau') \partial_{f} b^{\dagger}(\tau',z_{f}) G_{f}(\tau' - \tau) = 0 . \nn \label{RG_Equation_from_GR_Derivative} \eqa
Applying $\int_{0}^{\infty} d \tau$ into both terms, the resulting equation is nothing but Eq. (\ref{RG_Equation_from_GR}), which implies self-consistency of our formulation.

An idea is to compare Eq. (\ref{RG_Equation_from_GR}) with the renormalization group equation \cite{Bianchi:2001kw,deBoer:1999tgo,Verlinde:1999xm,AdS_CFT_Review}
\bqa && \gamma_{f}^{00} T_{00}^{f} + \beta_{f} \langle \mathcal{O}_{f} \rangle = 0 , \label{AdS/CFT_RG_Equation} \eqa
where an anomaly term does not arise in this Kondo problem. Here, either subscript or superscript $f$ denotes $z = z_{f}$. $\gamma_{f}^{00}$ is an induced metric and $T_{00}^{f}$ is an energy-momentum tensor at $z = z_{f}$. $\beta_{f}$ is a $\beta-$function of a coupling constant in the renormalization group analysis and $\langle \mathcal{O}_{f} \rangle$ is a vacuum expectation value of an observable $\mathcal{O}_{f}$ at $z = z_{f}$.

The $\beta-$function of the hybridization order parameter field is given by
\bqa && \beta_{f} = \partial_{f} b^{\dagger}(\tau,z_{f}) \nn && = - {J_{K}\over\pi\Lambda_b} \int_{0}^{\infty} d \tau' G_{c}(\tau-\tau') b^{\dagger}(\tau',z_{f}) G_{f}(\tau' - \tau) , \label{Beta_Function_RG} \eqa
where the last equality results from the IR boundary condition. The vacuum expectation value of the hybridization order parameter is given by
\bqa && \Big\langle \int \frac{d^{d} \bm{k}}{(2\pi)^{d}} c_{\sigma}^{\dagger}(\bm{k},\tau) f_{\sigma}(\tau) \Big\rangle \nn && = N \int_{0}^{\infty} d \tau' G_{c}(\tau-\tau') b(\tau',z_{f}) G_{f}(\tau'-\tau) . \label{Expectation_Value_RG} \eqa
As a result, we identify
\bqa && N \int_{0}^{\infty} d \tau b(\tau,z_{f}) \nn && \times \int_{0}^{\infty} d \tau' G_{c}(\tau-\tau') \partial_{f} b^{\dagger}(\tau',z_{f}) G_{f}(\tau'-\tau) \nonumber \eqa
with
\bqa && \int_{0}^{\infty} d \tau \beta_{f} \langle \mathcal{O}_{f} \rangle , \nonumber \eqa
where
\bqa && \mathcal{O}_{f} = \int \frac{d^{d} \bm{k}}{(2\pi)^{d}} c_{\sigma}^{\dagger}(\bm{k},\tau) f_{\sigma}(\tau) \nonumber \eqa
and $\tau \longleftrightarrow \tau'$.

The above correspondence gives rise to
\bqa && \int_{0}^{\infty} d \tau \gamma_{f}^{00} T_{00}^{f} \nn && = \frac{2 N S N_{F}}{v_{F} \Lambda_{c}^{2}} \int_{0}^{\infty} d \tau b(\tau,z_{f}) \partial_{\tau} b^{\dagger}(\tau,z_{f}) , \nonumber \eqa
or more strongly,
\bqa && \gamma_{f}^{00} T_{00}^{f} = \frac{2 N S N_{F}}{v_{F} \Lambda_{c}^{2}} b(\tau,z_{f}) \partial_{\tau} b^{\dagger}(\tau,z_{f}) , \label{RG_Energy_Momentum_Tensor_Part} \eqa
guaranteed by Eq. (\ref{RG_Equation_from_GR_Derivative}).

Substituting the IR boundary condition Eq. (\ref{IR_Boundary_Condition}) into Eq. (\ref{Effective_Free_Energy_Extra_Dimension}), the effective IR action further reduces to
\bqa && \mathcal{S}_{IR} = \int_{0}^{\infty} d \tau \Big\{ f_{\sigma}^{\dagger}(\tau) \partial_{\tau} f_{\sigma}(\tau) \nn && + \int_{0}^{\infty} d \tau' f_{\sigma}^{\dagger}(\tau) b(\tau,z_{f}) G_{c}(\tau-\tau') b^{\dagger}(\tau',z_{f}) f_{\sigma}(\tau') \Big\} , \nn \eqa
where the Lagrange multiplier field $\lambda$ is integrated out, giving rise to the number constraint $f_{\sigma}^{\dagger}(\tau) f_{\sigma}(\tau) = N S$. Then, it is straightforward to read the corresponding energy  as follows
\bqa T_{00}^{f} &=& \int_{0}^{\infty} d \tau' \Big\langle f_{\sigma}^{\dagger}(\tau) b(\tau,z_{f}) G_{c}(\tau-\tau') b^{\dagger}(\tau',z_{f}) f_{\sigma}(\tau') \Big\rangle \nn &=& N b(\tau,z_{f}) \int_{0}^{\infty} d \tau' G_{c}(\tau-\tau') b^{\dagger}(\tau',z_{f}) G_{f}(\tau' - \tau) , \nn \label{Metric_Ansatz} \eqa
which is nothing but the time component of the energy-momentum tensor.

Inserting the IR effective Hamiltonian [Eq. (\ref{Metric_Ansatz})] into the energy-momentum tensor part [Eq. (\ref{RG_Energy_Momentum_Tensor_Part})] of the renormalization group equation and resorting to the renormalization group equation (\ref{RG_Equation_from_GR}) with the IR boundary condition [Eq. (\ref{IR_Boundary_Condition})], we reach the following expression for the time component of the metric tensor
\begin{widetext}
\bqa && \gamma^{00}(\tau,z_{f}) = {J_{K}\over\pi\Lambda_b} \frac{\int_{0}^{\infty} d \tau' \int_{0}^{\infty} d \tau'' G_{c}(\tau-\tau') G_{c}(\tau'-\tau'') b^{\dagger}(\tau'',z_{f}) G_{f}(\tau'' - \tau') G_{f}(\tau' - \tau)}{\int_{0}^{\infty} d \tau' G_{c}(\tau-\tau') b^{\dagger}(\tau',z_{f}) G_{f}(\tau' - \tau)} . \label{Induced_Metric} \eqa
\end{widetext}

Inserting the non-perturbative solution Eq. (\ref{Nonperturbative_Solution}) for the hybridization order parameter into the above expression and performing the Fourier transformation into the frequency space, we obtain
\bqa && \gamma^{00}(z_{f}) = {J_{K}\over\pi\Lambda_b} \frac{1}{\beta} \sum_{i \omega} G_{c}(i \omega + C) G_{f}(i \omega) . \eqa
We already saw the right hand side in Eq. (\ref{Self_Consistent_Equation_C}). Using this result, we obtain
\bqa && \gamma^{00}(z_{f}) = - {N_{F} J_{K}\over\pi\Lambda_b} \ln \frac{\Lambda_{f}}{T_{K}} . \eqa
We recall
\bqa && T_{K} \approx \pi N_{F} \Big\{ \frac{C}{D} \cos \Big( \frac{C}{D} z_{f} \Big) + \sin \Big( \frac{C}{D} z_{f} \Big) \Big\}^{2} . \nonumber \eqa

The Ward identity in Eq. (\ref{AdS/CFT_RG_Equation}) represents the renormalization group flow of the boundary quantum field theory, which describes the relationship between physical quantities at the renormalization scale $z_f$. The effective metric $\gamma^{00}_f$ at the energy scale $z_f$ can be determined by solving this Ward identity if we know other quantities such as $T_{00}^f$, $\beta_f$, and $\left< {\cal O}_f \right>$, actually given by equations of motion and IR effective field theory. One thing we should note is that the renormalization group flow must satisfy the Ward identity regardless of the renormalization scale $z_f$. This fact implies that we can see how $\gamma^{00}_f$ changes as the renormalization scale runs from the UV to IR scale. From the holographic point of view, the radial coordinate of the dual geometry can be matched to the energy scale of the dual field theory. More precisely, when the boundary of the dual geometry is located at $z_f$, the dual field theory is defined at the renormalization scale $z_f$ and the effective metric $\gamma^{00}_f$ corresponds to the metric of the boundary spacetime. Therefore, the renormalization group flow of the dual field theory can be understood by changing the boundary position on the dual geometry side \cite{deBoer:1999tgo,Verlinde:1999xm}.

\subsection{How to find bulk metric}

The next question is how to find the bulk metric consistent with the boundary metric given by the Hamilton-Jacobi formulation. Frankly speaking, we do not know how to figure out the bulk metric in the Hamilton-Jacobi formulation. Here, we propose another way and discuss consistency with the induced boundary matric.

We recall an effective bulk action
\bqa && \mathcal{S}_{bulk} = \int_{0}^{\infty} d z \int_{0}^{\infty} d \tau \bigg\{ \frac{\pi N\Lambda_b}{J_{K} M^{2}} \Big( \partial_{z} b^{\dagger}(\tau,z) \Big) \Big( \partial_{z} b(\tau,z) \Big) \nn && + \frac{2 N S N_{F}}{v_{F} \Lambda_{c}^{2}} b(\tau,z) \partial_{\tau} b^{\dagger}(\tau,z) \bigg\} , \eqa
which describes the evolution of the hybridization order parameter as a function of an energy scale. A mass parameter $M^{2}$ has been introduced, the reason of which will be discussed below.

Our strategy is to compare this effective bulk action with the canonical form of a spin $0$ bosonic field theory on a curved spacetime and to read out both metric components of $g_{\tau\tau}$ and $g_{zz}$. Unfortunately, our effective action contains a linear time-derivative term. Since the canonical form of the spin $0$ bosonic field theory on a curved spacetime does not take the linear time-derivative term, we cannot compare our effective action with the canonical form of general covariance. In this respect we consider a reformulation with UV completion for this non-relativistic effective action. We consider the following effective action
\bqa && \mathcal{S}_{eff} = \int_{0}^{\infty} d z \int_{0}^{\infty} d \tau \bigg\{ \frac{2\pi N\Lambda_b}{J_{K}} \Big( \partial_{z} b^{\dagger}(\tau,z) \Big) \Big( \partial_{z} b(\tau,z) \Big) \nn && + \Big( \frac{2 N S N_{F}}{v_{F} \Lambda_{c}^{2}} \Big)^{2} \Big( \partial_{\tau} b(\tau,z) \Big) \Big( \partial_{\tau} b^{\dagger}(\tau,z) \Big) + M^{2} b(\tau,z) b^{\dagger}(\tau,z) \bigg\} , \nn \eqa
where the second-order time derivative term has been taken into account with the introduction of a mass term. Considering the $M^{2} \rightarrow \infty$ limit, keeping only the positive-energy part, and throwing away the huge mass term of the boson field, we return to our original non-relativistic effective field theory.

It is natural to consider the following canonical form
\bqa && \mathcal{S}_{eff} = \int_{0}^{\infty} d z \int_{0}^{\infty} d \tau \sqrt{g_{\tau\tau} g_{zz}} \nn && \bigg\{ \frac{2\pi N\Lambda_b}{J_{K}} g^{zz} \Big( \partial_{z} b^{\dagger}(\tau,z) \Big) \Big( \partial_{z} b(\tau,z) \Big) \nn && + \Big( \frac{2 N S N_{F}}{v_{F} \Lambda_{c}^{2}} \Big)^{2} g^{\tau\tau} \Big( \partial_{\tau} b(\tau,z) \Big) \Big( \partial_{\tau} b^{\dagger}(\tau,z) \Big) \nn && + M^{2} b(\tau,z) b^{\dagger}(\tau,z) \bigg\} , \eqa
where metric components have been introduced. It is straightforward to read out both metric components as follows
\bqa && g_{\tau\tau} = g_{\tau\tau}^{*} = 1 , ~~~~~ g_{zz} = g_{zz}^{*} = 1 . \eqa
Then, the question is how this Euclidean metric can be consistent with the induced metric found from the Callan-Symanzik equation in the previous subsection. Although we cannot prove our speculation, we suspect that a general coordinate transformation leads the boundary metric to be the Euclidean in the bulk.

\subsection{From strongly coupled quantum field theory to Einstein's gravity formulation}

Now, we discuss how to derive Einstein's theory of general relativity as a dual description of a strongly coupled quantum field theory. We start from an effective field theory
\bqa && Z = \int D b(\tau,z) D g_{\tau\tau}(\tau,z) D g_{zz}(\tau,z) \nn && \delta\Big(g_{\tau\tau}(\tau,z) - g_{\tau\tau}^{*}[b(\tau,z)]\Big) \delta\Big(g_{zz}(\tau,z) - g_{zz}^{*}[b(\tau,z)]\Big) \nn && \exp\bigg[ - \int_{0}^{\infty} d z \int_{0}^{\infty} d \tau \sqrt{g_{\tau\tau}(\tau,z) g_{zz}(\tau,z)} \nn && \bigg\{ \frac{2\pi N\Lambda_b}{J_{K}} g^{zz}(\tau,z) \Big( \partial_{z} b^{\dagger}(\tau,z) \Big) \Big( \partial_{z} b(\tau,z) \Big) \nn && + \Big( \frac{2 N S N_{F}}{v_{F} \Lambda_{c}^{2}} \Big)^{2} g^{\tau\tau}(\tau,z) \Big( \partial_{\tau} b(\tau,z) \Big) \Big( \partial_{\tau} b^{\dagger}(\tau,z) \Big) \nn && + M^{2} b(\tau,z) b^{\dagger}(\tau,z) \bigg\} \bigg] . \eqa
Both metric components of $g_{\tau\tau}^{*}[b(\tau,z)]$ and $g_{zz}^{*}[b(\tau,z)]$ are found from the comparison between the effective Landau-Ginzburg field theory with an extra dimension and the canonical form of the corresponding field theory as discussed in the previous subsection. In the Kondo problem (perfect screening \cite{Kondo_Textbook}), they do not depend on the hybridization order parameter, given by the Euclidean metric. In the present subsection, we put our discussion on a more general setup.

Utilizing the Lagrange multiplier field, we can exponentiate the delta-function constraint and obtain the following expression
\bqa && Z = \int D b(\tau,z) D g_{\tau\tau}(\tau,z) D g_{zz}(\tau,z) D T_{\tau\tau}(\tau,z) \nn && D T_{zz}(\tau,z) \exp\bigg[ - \int_{0}^{\infty} d z \int_{0}^{\infty} d \tau \sqrt{g_{\tau\tau}(\tau,z) g_{zz}(\tau,z)} \nn && \bigg\{ \frac{2\pi N\Lambda_b}{J_{K}} g^{zz}(\tau,z) \Big( \partial_{z} b^{\dagger}(\tau,z) \Big) \Big( \partial_{z} b(\tau,z) \Big) \nn && + \Big( \frac{2 N S N_{F}}{v_{F} \Lambda_{c}^{2}} \Big)^{2} g^{\tau\tau}(\tau,z) \Big( \partial_{\tau} b(\tau,z) \Big) \Big( \partial_{\tau} b^{\dagger}(\tau,z) \Big) \nn && + M^{2} b(\tau,z) b^{\dagger}(\tau,z) \nn && - i T_{\tau\tau}(\tau,z) \Big(g^{\tau\tau}(\tau,z) - g^{\tau\tau *}[b(\tau,z)]\Big) \nn && - i T_{zz}(\tau,z) \Big(g^{zz}(\tau,z) - g^{zz *}[b(\tau,z)]\Big) \bigg\} \bigg] , \eqa
where $T_{\tau\tau}(\tau,z)$ and $T_{zz}(\tau,z)$ are Lagrange multiplier fields. Integrating over metric fields $g^{\tau\tau}(\tau,z)$ and $g^{zz}(\tau,z)$, we find
\bqa && T_{\tau\tau}(\tau,z) = \Big( \frac{2 N S N_{F}}{v_{F} \Lambda_{c}^{2}} \Big)^{2} \Big( \partial_{\tau} b(\tau,z) \Big) \Big( \partial_{\tau} b^{\dagger}(\tau,z) \Big) , \nn && T_{zz}(\tau,z) = \frac{2\pi N\Lambda_b}{J_{K}} \Big( \partial_{z} b^{\dagger}(\tau,z) \Big) \Big( \partial_{z} b(\tau,z) \Big) . \eqa
In this respect these Lagrange multiplier fields may be identified with energy-momentum tensors as collective fields.

The last step is to integrate out the order parameter field. Formally, such a procedure is expected to result in
\bqa && Z = \int D g_{\tau\tau}(\tau,z) D g_{zz}(\tau,z) T_{\tau\tau}(\tau,z) T_{zz}(\tau,z) \nn && \exp\bigg[ - \int_{0}^{\infty} d z \int_{0}^{\infty} d \tau \sqrt{g_{\tau\tau}(\tau,z) g_{zz}(\tau,z)} \nn && \bigg\{ \mathcal{L}_{eff}[g^{\mu\nu}(\tau,z),T_{\mu\mu}(\tau,z)] - i T_{\tau\tau}(\tau,z) g^{\tau\tau}(\tau,z) \nn && - i T_{zz}(\tau,z) g^{zz}(\tau,z) \bigg\} \bigg] . \eqa
Here, $\mathcal{L}_{eff}[g^{\mu\nu}(\tau,z),T_{\mu\mu}(\tau,z)]$ would contain the Einstein-Hilbert action in the presence of energy-momentum tensor collective fields.

\section{Summary}

Our emergent holographic description with an extra dimension turned out to be a Landau-Ginzburg theory of local order parameter fields, renormalized by quantum corrections and described by Wilsonian renormalization group transformations. The extra dimension is identified with the number of iterations of renormalization group transformations. Here, we derived an effective field theory of a hybridization order parameter field for the Kondo effect, living on the spacetime with an extra dimension given by Wilsonian renormalization group transformations. In order to reveal the physics of this extra dimension, we took the limit of $z_{f} = d z \longrightarrow 0$ and showed that the resulting equation of motion for the hybridization order parameter is reduced into that of a mean-field theory with leading quantum corrections in the $1/N$ expansion of the field theoretical approach, where $N$ is spin degeneracy. This demonstration not only serves as a solid background of the present approach but also implies a non-perturbative framework with full quantum corrections in the $z_{f} \longrightarrow \infty$ limit.

It was rather unexpected to observe that the locality of the effective field theory is preserved in recursive Wilsonian renormalization group transformations. It is natural to expect the appearance of nonlocal interactions in the Wilsonian renormalization group transformation. Indeed, nonlocal effective interactions arise generically. However, we find that such effective interactions depend on the UV cutoff, where nonlocal correlations vanish in the infinite limit of the UV cutoff. In other words, nonlocal correlations can be expressed as local terms through the gradient expansion when the UV cutoff is large enough.

Based on this effective field theory, we obtained a classical equation of motion in the large $N$ limit, which describes how the hybridization order parameter field evolves as a function of the energy scale, identified with the extra dimension. In addition to this equation of motion, UV and IR boundary conditions were naturally constructed to fix the resulting configuration of the hybridization order parameter field in the extra dimension unambiguously. Solving these coupled equations of motion, we found the hybridization order parameter field as a function of both time and the energy scale of the extra dimension. It turned out that this solution reproduces the local Fermi-liquid fixed point in the absence of any phase transitions at finite temperatures. In addition, we investigated thermodynamic properties in more details, and compared them with those of the Bethe ansatz method. As a result, we could show that our non-perturbative theoretical framework is consistent with the essentially exact solution, which describes the crossover behavior from the decoupled local moment fixed point to the local Fermi-liquid fixed point quite successfully. This reflects the non-perturbative nature of our formulation.

We also discussed how to extract out an emergent metric structure from this effective field theory. Resorting to the Hamilton-Jacobi formulation derived from our effective free-energy functional, we could read the $\beta-$function for the evolution of the hybridization order parameter and its vacuum expectation value. It turned out that the Callan-Symanzik equation for the effective free energy just reformulates our coupled equations of motion, i.e., the bulk equation of motion for the order parameter field and the UV and IR boundary conditions, which confirms the internal consistency of our formulation. Based on this reformulation, we could find an induced metric.

Finally, we compared our recursive Wilsonian renormalization group structure with the Wilson's numerical renormalization group method. In particular, we argued that recursive Wilsonian renormalization group transformations are in parallel with the recursive exact diagonalization procedure of the numerical renormalization group structure. As a result, we identified the size $N$ of the numerical renormalization group transformation with the emergent extra dimension $z$ of our recursive renormalization group transformation. This implies that the theoretical framework of recursive renormalization group transformations can be applied to various problems of strongly correlated systems as the numerical renormalization group method does. In particular, it would be interesting to apply the present theoretical framework to various Kondo problems such as the case of overscreening and the presence of competing Hund interactions. Furthermore, the correspondence between the recursive Wilsonian renormalization group structure and the Wilson's numerical renormalization group method should be more clarified based on the analytic construction for recursion equations of renormalization group flows in the numerical renormalization group structure.

\begin{acknowledgments}
This study was supported by the Ministry of Education, Science, and Technology (No. NRF-2015R1C1A1A01051629, No. 2011-0030046, and No. NRF-2013R1A1A2A10057490) of the National Research Foundation of Korea (NRF) and by TJ Park Science Fellowship of the POSCO TJ Park Foundation. This work was also supported by the POSTECH Basic Science Research Institute Grant (2016). C. Park was supported by the Basic Science Research Program through the National Research Foundation of Korea funded by the Ministry of Education (NRF-2016R1D1A1B03932371) and by the Korea Ministry of Education, Science and Technology, Gyeongsangbuk-Do, and Pohang City. S.B.C was also supported by IBS-R009-Y1. We would like to appreciate fruitful discussions in the APCTP Focus program ``Lecture series on Beyond Landau Fermi liquid and BCS superconductivity near quantum criticality" in 2016. K.-S. Kim appreciates fruitful discussions with G.-S. Jeon for the numerical renormalization group method. K.-S. Kim also thanks M. Kiselev for his suggestion on how to deal with the single occupancy constraint exactly although not applied in the present study.
\end{acknowledgments}


\begin{thebibliography}{9}
\bibitem{Comment_on_Frequency_Domain} We point out that the integration over electron frequencies can go all the way to values $\sim v_F \Lambda_c$. This does not cause any problems in the renormalization group analysis. Recall that both holons and spinons are defined in the low-frequency domain. Although the electron frequency $\omega$ covers all values until $v_{F} \Lambda_{c}$, these contributions cannot allow such holons and spinons to exist in the low-frequency domain. In this respect only electron frequencies to satisfy the condition of $v_{F} \Lambda_{c} \gg \Lambda_{f}, ~ \Lambda_{b} \geq |\omega|$ contribute to this renormalization. Actually, the condition of $v_{F} \Lambda_{c} \gg \Lambda_{f}, ~ \Lambda_{b}$ is physical for the Kondo effect, naturally imposed in the renormalization group analysis.
\end{thebibliography}
\end{document}